\documentclass{ar-1col-S2O}

\usepackage[comma]{natbib}
\usepackage{url}
\usepackage{graphicx}
\usepackage{subcaption}
\usepackage{soul}

\jname{Ann. Rev. Astronomy and Astrophysics}
\jvol{62}
\jyear{2024}
\doi{10.1146/((article doi))}

\newcommand\sun{\odot}
\newcommand{\solar}{\ifmmode_{\sun}\;\else$_{\sun}\;$\fi}
\newcommand{\HI}{H$\,${\sc i}}
\newcommand{\HII}{H$\,${\sc ii}}
\newcommand{\ha}{H$\alpha$}

\newcommand{\unithi}{M\solar pc$^{-2}$}
\newcommand{\kms}{km s$^{-1}$}
\newcommand{\arcsec}{\ifmmode^{\prime\prime}\;\else$^{\prime\prime}\;$\fi}
\newcommand{\arcmin}{\ifmmode^{\prime}\;\else$^{\prime}\;$\fi}
\newcommand*\aap{A\&A}

\newcommand*\aapr{A\&A~Rev.}

\newcommand*\aj{AJ}
\newcommand*\apj{ApJ}
\newcommand*\apjl{ApJ}

\newcommand*\apjs{ApJS}

\newcommand*\araa{ARA\&A}
\newcommand*\mnras{MNRAS}

\newcommand*\nar{New A Rev.}
\newcommand*\nat{Nature}

\newcommand*\pasj{PASJ}
\newcommand*\pasp{PASP}

\newcommand{\hers}{{\it  Herschel}}

\newcommand{\iso}{{\it  ISO}}

\newcommand{\spitz}{{\it   Spitzer}}
\newcommand{\iras}{{\it  IRAS}}

\newcommand{\lfir}{L$_{FIR}$}

\newcommand{\mic}{$\mu$m}

\newcommand{\zsol}{$Z_\odot$}

\newcommand{\smallsub}[1]{{\mbox{{\tiny #1}}}}

\newcommand{\Av}{A$_\smallsub{V}$}

\newcommand{\hmol}{H$_2$}
\newcommand{\hi}{H$\,${~\sc i}}

\newcommand{\cii}{[C$\,${~\sc ii}]}
\newcommand{\ci}{[C$\,${~\sc i}]}

\newcommand{\ciiline}{\cii$\lambda 158$\mic}

\newcommand{\mhmol}{M$_\smallsub{H$_2$}$}
\newcommand{\ciico}{L$_\mathrm{[C\,{\sc II}]}$/L$_\mathrm{CO(1-0)}$}

\begin{document}

\markboth{Hunter et al.}{ISM in Dwarfs}

\title{The Interstellar Medium in Dwarf Irregular Galaxies}

\author{Deidre A.\ Hunter,$^1$ Bruce G.\ Elmegreen,$^2$ and Suzanne C.\ Madden$^3$
\affil{$^1$Lowell Observatory, Flagstaff, USA, 86001; email: dah@lowell.edu; ORCID 0000-0002-3322-9798}
\affil{$^2$IBM Research, T.J.\ Watson Research Center, 1101 Kitchawan Road, Yorktown Heights, NY 10598; email: belmegreen@gmail.com; ORCID 0000-0002-1723-6330}
\affil{$^3$Universit\'{e} Paris Cit\'{e}, Universit\'{e} Paris-Saclay, CEA, CNRS, AIM, F-91191, Gif-sur-Yvette Cedex, France; email: suzanne.madden@cea.fr; ORCID 0000-0003-3229-2899}}

\begin{abstract}
Dwarf irregulars (dIrrs) are among the most common type of galaxy in the Universe. They typically have gas-rich, low surface-brightness, metal-poor, and relatively-thick disks. 
Here we summarize the current state of our knowledge of the interstellar medium (ISM), including atomic, molecular and ionized gas, along with their dust properties and metals.
We also discuss star formation feedback, 
gas accretion, and mergers with other dwarfs
that connect the ISM to the circumgalactic and intergalactic media.
We highlight one of the most persistent mysteries: the nature of pervasive gas that is yet undetected as either molecular or cold hydrogen, the ``dark gas''.  
Here are a few highlights:
\begin{itemize}
\item Significant quantities of \HI\ are in far-outer gas disks.
\item Cold \HI\ in dIrrs would be molecular in the
Milky Way, making the\\
chemical properties of star-forming clouds significantly different.
\item Stellar feedback has a much larger impact in dIrrs than in \\
spiral galaxies.
\item The escape fraction of ionizing photons is significant, making dIrrs \\
a plausible source for reionization in the early Universe. 
\item Observations suggest a significantly higher abundance of hydrogen \\
(H$_2$ or cold \HI)
associated with CO in star-forming regions than that \\
traced by the CO alone.
\end{itemize}
\end{abstract}

\maketitle

\tableofcontents

\section{INTRODUCTION} \label{sect-intro}

The interstellar medium (ISM) is a dynamic part of galactic ecosystems.
Baryons are cycled from the ISM into stars through star formation processes 
and are returned to the ISM, circumgalactic medium (CGM), and even intergalactic medium (IGM)
through feedback from stellar winds and explosions. With stellar feedback in the weak gravitational potential of a dIrr, 
the products of stellar nucleosynthesis are only partially redistributed throughout the ISM; significant quantities escape into the halo, leaving the disk relatively metal-poor. Overall the ISM is complex, dynamic, and 
multi-faceted.

\begin{figure}[t!]
\begin{subfigure}{.5\textwidth}
    \centering
    \includegraphics[width=2.25in]{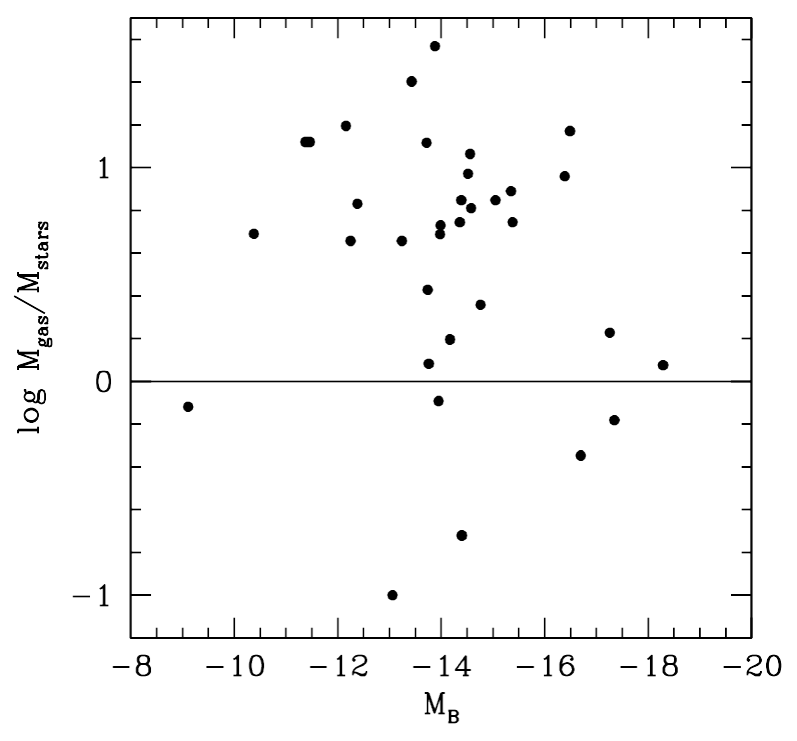}
\end{subfigure}
\begin{subfigure}{.5\textwidth}
    \centering
    \includegraphics[width=2.25in]{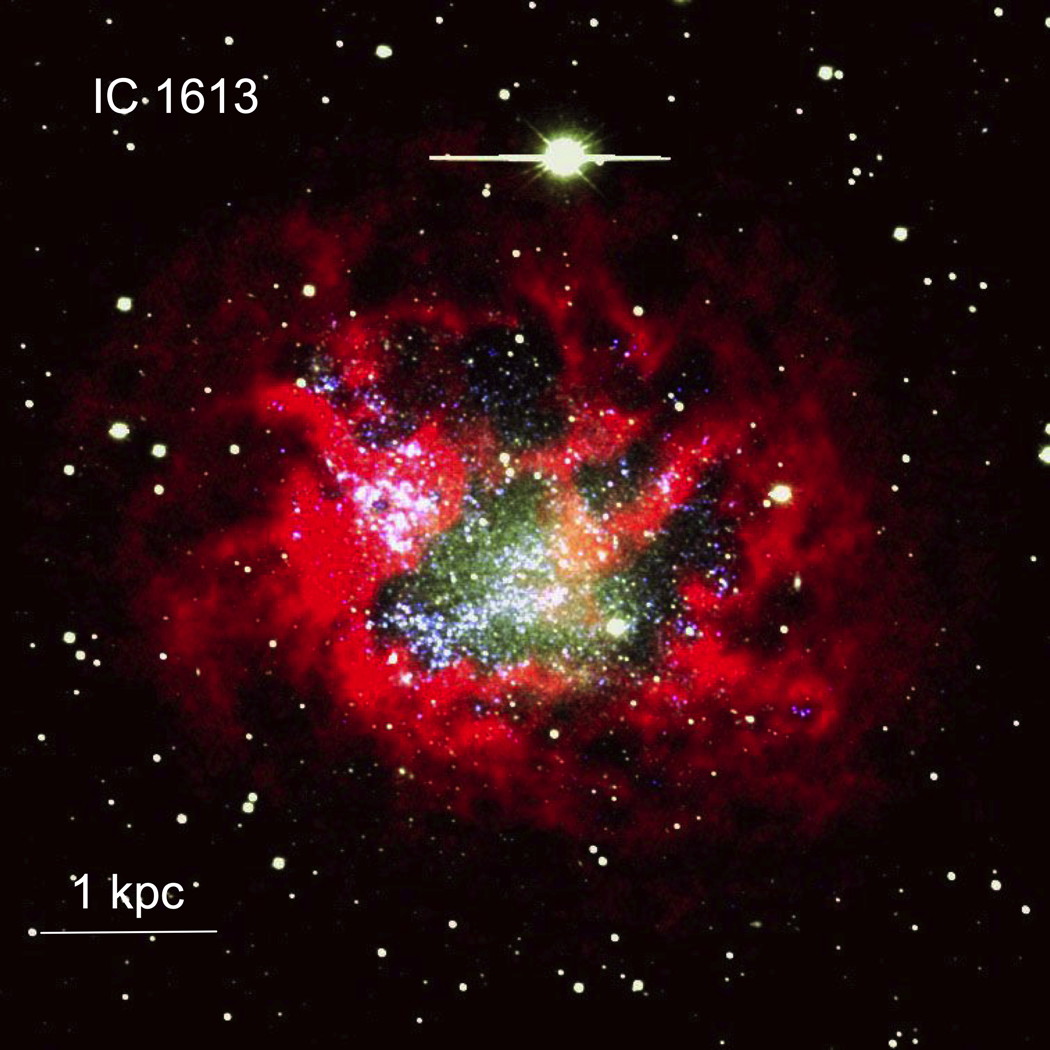}
\end{subfigure}
\caption{
\textit{Left:} Logarithm of the ratio of atomic gas mass (\HI$+$He) to stellar mass as a function of galaxy absolute $B$ magnitude. 
These data are 
from \citet{Zhang:12b} for a subsample of the LITTLE THINGS nearby dIrr galaxies 
\citep[see][]{Hunter:12}.
The horizontal line marks a ratio of 1.
\textit{Right:}
False-color image of IC 1613, a typical dIrr in the Local Group. 
Blue is far-ultraviolet, green is $V$-band, and red is atomic hydrogen gas.
From \citet{Hunter:12}, reproduced by permission of the AAS. We thank Lauren Hill for producing the image on the right.} 
\label{fig-intro}
\end{figure}

In this review, we concentrate on the ISM of dwarf Irregular (dIrr) galaxies. 
dIrrs are tiny but they are not just small versions of spiral galaxies. 
dIrrs are, in fact, dominated by their ISM
(see, for example, 
\textbf{Figure \ref{fig-intro}}).
Their ISMs are primarily atomic hydrogen and helium but the inner regions could also contain 
substantial amounts of molecular hydrogen (H$_2$), and they most often have star formation. 
Dwarf spheroidal galaxies (dSph), on the other hand, are primarily devoid of an ISM, believed to have been stripped.
Thus, we will not discuss dSphs.

\begin{marginnote}[]
\entry{ISM}{The interstellar medium is the gas and dust mixed with the stars of a galaxy.}
\entry{CGM}{The circumgalactic medium is the gas in the outer halo of a galaxy.}
\entry{IGM}{The intergalactic medium is the gas beyond and between galaxies.}
\end{marginnote}

Dwarf galaxies were first studied in detail by
\citet{reaves56} in a survey for them in the Virgo Cluster. Unfortunately, the class was never truly defined; Reaves refers to them simply as ``underluminous.'' In subsequent decades some researchers defined samples based on luminosity, some on surface brightness, and some on baryonic mass. Furthermore, 50 years ago the dwarfs were labeled as Irr-I or Im (Magellanic-type irregulars) or Irr-II (amorphous). However, in recent decades, as dwarfs became better known, those terms have been dropped and researchers rarely define ``dwarf''. Here we will define what we mean by ``dIrr'' by two galaxies that represent the extremes to us. 

Dwarf irregulars are generally faint, small, low in total mass, low in metallicity, low in dust, and blue in optical colors. 
However, as a class they exhibit a range in these properties
\citep[see lists of dIrrs in][]{Mateo:98,Hunter:12}.
In the Local Group, for example, Leo T has an absolute $V$ magnitude $M_V$ of $-7$ \citep{deJong:08},
a total baryonic mass of only $6\times10^5$ $M$\solar, a half-light radius of 170 pc, and a metallicity 0.025$Z$\solar 
\citep{Irwin:07,Ryan-Weber:08,Adams:18}.
The Large Magellanic Cloud (LMC), on the other hand, has $M_V=-18$ \citep{Macri:06},
a baryonic mass of $3\times10^9$ $M$\solar \citep{Besla:15}, a radius at 25 mag arcsec$^{-2}$ in the $B$-passband of 4.9 kpc \citep{rc3}, and a metallicity of 0.5$Z$\solar.
At another extreme, J0139$+$4328 is a dwarf with $8\times10^7$ $M$\solar of gas but no detectable  stars 
\citep[$M_{\rm stars} < 7\times10^5$ $M$\solar,][]{Xu:23}.
A typical nearby dIrr is shown in 
\textbf{Figure \ref{fig-intro}} and others are shown in figures throughout the text.

\begin{marginnote}[]
\entry{dIrr}{Dwarf irregular galaxies are small and intrinsically chaotic in appearance. 
They lack spiral arms and bulges.
}
\entry{dSph}{Dwarf spheroidal galaxies have a smooth, elliptical appearance.}
\entry{BCD}{Blue Compact Dwarfs are centrally concentrated, high surface brightness, and actively forming stars.}
\end{marginnote}

Although they are tiny compared to spiral and elliptical galaxies, dwarf systems have had an
outsized cosmological impact. Simulations of galaxy formation show spirals forming from
merging small dark-matter halos, which are essentially dwarf galaxies \citep[see][]{tosi03}. 
The remnants of
such accretions are sometimes visible today as stellar streams in galaxy halos \citep[e.g.,][]{ibata94}. 
Dwarfs are also likely to have formed the first stars  \citep[see review in][]{bromm04}.
Lyman-alpha emitters at high redshift have low masses like dwarfs \citep{ouchi20} and
may have some properties in common with local dIrrs. Their relatively high escape fraction for Lyman continuum radiation suggests they contributed to the 
reionization of the Universe \citep[e.g.,][]{asada23}.

Below we discuss the various components of the ISM of dIrrs, the processes that take place there, 
exchange of gas from the CGM and IGM, and outstanding questions.
Major surveys related to this topic are listed in \textbf {Table \ref{tab1}}.
Related reviews include those on the cold ISM in galaxies \citep{Saintongue:22}, dust in galaxies \citep{galliano18}, 
the ISM in dwarfs \citep{Henkel:22}, chemical properties of dwarf galaxies \citep{Annibali:22}, 
warm and hot diffuse gas in dwarfs \citep{Bomans:01},
ultra-faint dwarfs \citep{simon19}, 
the structure and evolution of irregular galaxies \citep{GH84},
and star formation in dwarfs \citep{Hunter:08,Bolatto:19}.

\section{ATOMIC GAS} \label{sect-atomicgas}

\subsection{Large-scale Distribution} \label{sect-atomic-largescale}

Atomic hydrogen gas densities in dIrr galaxies are generally lower than in giant spirals.
Surface densities in the central regions have peaks of order 1-8 \unithi, which are comparable
to those in the main disk of spirals, 5-10 \unithi\ \citep{Brinks:16,Hunter:21b}. 
However, spiral galaxies have much higher surface densities closer to the center, while dIrrs have lower  surface densities everywhere else.
Fits of a Sersic function ($I(R) = I_0 e^{-(R/R_0)^{1/n}}$)
to the \HI\ radial ($R$) profiles of dIrrs find $n\sim0.2-1$, where an $n=1$ is exponential; $n<1$ has a slightly flatter central profile and a faster fall-off outside  \citep{Hunter:21b}.

{\bf Figure \ref{fig-intro}} shows that most
dIrrs are gas-rich compared to their stellar mass. 
The gas associated with galaxies usually extends further than the stars, but it may extend much further than previously
realized. Deep \HI\ surveys such as MHONGOOSE, with column density sensitivities better than $10^{18}$ cm$^{-2}$,
find up to 25\% of a galaxy’s gas in the far outer regions at very low column densities
 \citep[see, for example,][]{Sardone:21}.
It is not clear what role this outer disk gas
plays in galactic evolution.

Gas in all but the smallest dIrrs exhibits regular circular rotation, often with a near-linear rise in rotation speed away
from the center 
\citep{Lo:93a,Lo:93b,Oh:15,Iorio:17}. 
Observations of the \HI\ kinematics of the low-mass end of dIrrs
show the difficulties in fitting a rotation curve
to these galaxies \citep[e.g.][]{Oh:15,SHIELD:16}.
Simulations also suggest that at the lowest masses ($<10^8$ $M$\solar), dIrrs should be dispersion-dominated 
because of heating by the cosmic UV background, which suppresses high angular momentum accretion, and high turbulence from stellar feedback \citep{El-Badry:18}.
For those galaxies with regular rotation, the ISM does not undergo the levels of shear that occur in spirals with flat rotation curves, and this low shear means that ISM structures, such as shells and holes, can last longer than in the highly sheared regime of spiral disks.

\subsection{Structures} \label{sect-atomic-structures}

The ISM of dIrr galaxies is not smoothly distributed. It contains clouds, holes, and shells. There is also flaring of the \HI\ disk and turbulence throughout the gas. Here we discuss the various structures in the ISM and the relationship of the ISM to star formation and the stellar disk.

\subsubsection{Clouds} \label{sect-atomic-clouds}

Atomic gas consists of cool and dense clouds, often in filaments, and lower-density, warmer gas between them. 
Power spectra of \HI\ emission from the SMC and LMC actually show a hierarchy of structures, so the cloud-intercloud model is oversimplified \citep{Stanimirovic:99,Elmegreen:01}.
Dense regions that might be identified as discrete clouds several 100 pc in size were identified in 9 dIrrs by \citet{Lo:93a,Lo:93b}.
\citet{Kim:07} identified 468 \HI\ clouds with masses of $10^3$ to $10^5$ $M$\solar in the LMC
and \citet{Hunter:19} identified 814 \HI\ clouds with masses of $10^3$ to $10^7$ $M$\solar in 40 nearby dIrrs.
In the Hunter et al.\ study, the clouds constitute 2\% to 53\% of the total \HI\ mass of the host galaxy.
Many of these clouds are in the outer regions, beyond one disk scale length, and
not all of them are self-gravitating unless they contain significant amounts of molecular or cold atomic gas.

\begin{marginnote}[]
\entry{SMC and LMC}{The Small and Large Magellanic Clouds are two dIrrs that are gravitationally interacting 
with each other and with the Milky Way. The LMC is at the massive end of
dwarfs, and the SMC has been significantly disturbed by the interaction.}
\end{marginnote}

\subsubsection{Holes, Shells, and Outflows} \label{sect-atomic-holes}

The ISM of dIrrs is especially characterized by holes, giving them a high porosity 
\citep{Dimaratos:15,Cormier:19}.
In 41 gas-rich dIrrs from LITTLE THINGS, 
\citet{Pokhrel:20} identified 306 holes in \HI\ ranging in diameter from the resolution
limit of 40 pc up to 2.3 kpc with expansion speeds up to
30 km s$^{-1}$. The \HI\ surface porosity 
(ratio of total area covered by holes to total area covered by \HI\ out to a column density of
$5\times10^{19}$ atoms cm$^{-2}$) is as high as 15\%. 
In another sample of nearby spirals and dwarfs, \citet{Bagetakos:11} detected more than 1000 holes
in 20 galaxies with sizes of 0.1 kpc to 2 kpc and ages of 3 Myr to  150 Myr; 23\% of these fell beyond the optical radius. Relatively low shear, reduced disk gravity, and a relatively large scale height in dIrrs enables feedback-driven shells to grow larger compared to the disk scale length than in spirals \citep{Brinks:02}.

Giant shells in dwarf galaxies often contain secondary star formation triggered by the compression of gas in the surrounding region. Examples are in 
IC 2574 \citep{walter98,egorov14}, 
DDO 47 \citep{Walter:01},
Haro 14 \citep{cairos17a},
Tololo 1937-423 \citep{cairos17b},
Holmberg I \citep{ott01,egorov18}, 
and
Holmberg II \citep[][see {\bf Figure \ref{fig-egorov17}} below]{puche92,egorov17}.
\citep[See][for a solar neighborhood example.]{zucker:22}
Simulations of this process are in \cite{kawata14} and \cite{lahen19}. 
Constellation III in the LMC is a spectacular region that formed stars 12--16 Myrs ago and blew a 1.4-kpc 
hole in the ISM. Stars have formed in the surrounding shell over the past 6 Myrs \citep{Dolphin:98}.
Some holes in dIrrs are large enough to dominate the entire galaxy. In DDO 88 
there is a 2-kpc hole centered on the optical galaxy with a radius of half of the optical size of the galaxy.
The shell surrounding the hole contains 30\% of the total \HI\  \citep{Simpson:05}. 
The impact of giant shells on total star formation is not known. Only a small fraction of the star formation ($<$15\%)
is in obvious shells \citep{Bolds:20}, and the star formation rate (SFR) does not correlate with the porosity \citep{Pokhrel:20}.

Beyond the immediate shells with triggered star formation, the porosity caused by holes
allows far-ultraviolet (FUV) photons from young stars to travel throughout a galaxy, heating the ISM 
and inhibiting  more star formation  \citep{Silk:97,forbes16}. 
Another consequence of this porosity is that star formation moves
around the galaxy on long time and spatial scales
as regions are hollowed out by stellar feedback and take time to fill back in.
Combined with relatively rapid motions from feedback and aggravation from outside, the effects can be severe, as in DDO 187 and NGC 3738, which have
a blue star-forming half and a redder half with only older stars \citep{Hunter:18}.
Heating and dynamical perturbations can also lead to ``gaspy'' star formation  \citep{Lee:07}
where the star formation rate increases and decreases as conditions periodically favor and then
inhibit star formation on a large scale.
This can even lead to a
reshuffling of the \HI\ relative to the stars
where the gas becomes more centrally concentrated \citep{Simpson:00} or of the stars themselves as stellar feedback forces radial migration of stars outward \citep{elbadry16}.


Holes are generally believed to be formed by feedback from massive stars. 
These stars dump copious quantities
of mechanical energy into the ISM from stellar winds and supernovae  \citep{TT:88,Vorobyov:05}. 
However, the story may sometimes be more complicated than holes resulting from a single star formation event.
Multi-age stellar populations are inside the \HI\ holes in Holmberg II, suggesting feedback over a period
of 100 Myrs \citep{Weisz:09}.
\citet{Kepley:07} suggested that the central \HI\ depression in  WLM could have formed by feedback over the past 130 Myr.
Multiple generations including triggered star formation were found to enlarge the holes in 5 dIrrs by \citet{Warren:11}. 
Starbursts occasionally last for 500 Myr \citep{mcquinn10}.

\begin{figure}[t!]
\begin{subfigure}{.5\textwidth}
    \centering
    \includegraphics[width=2.5in]{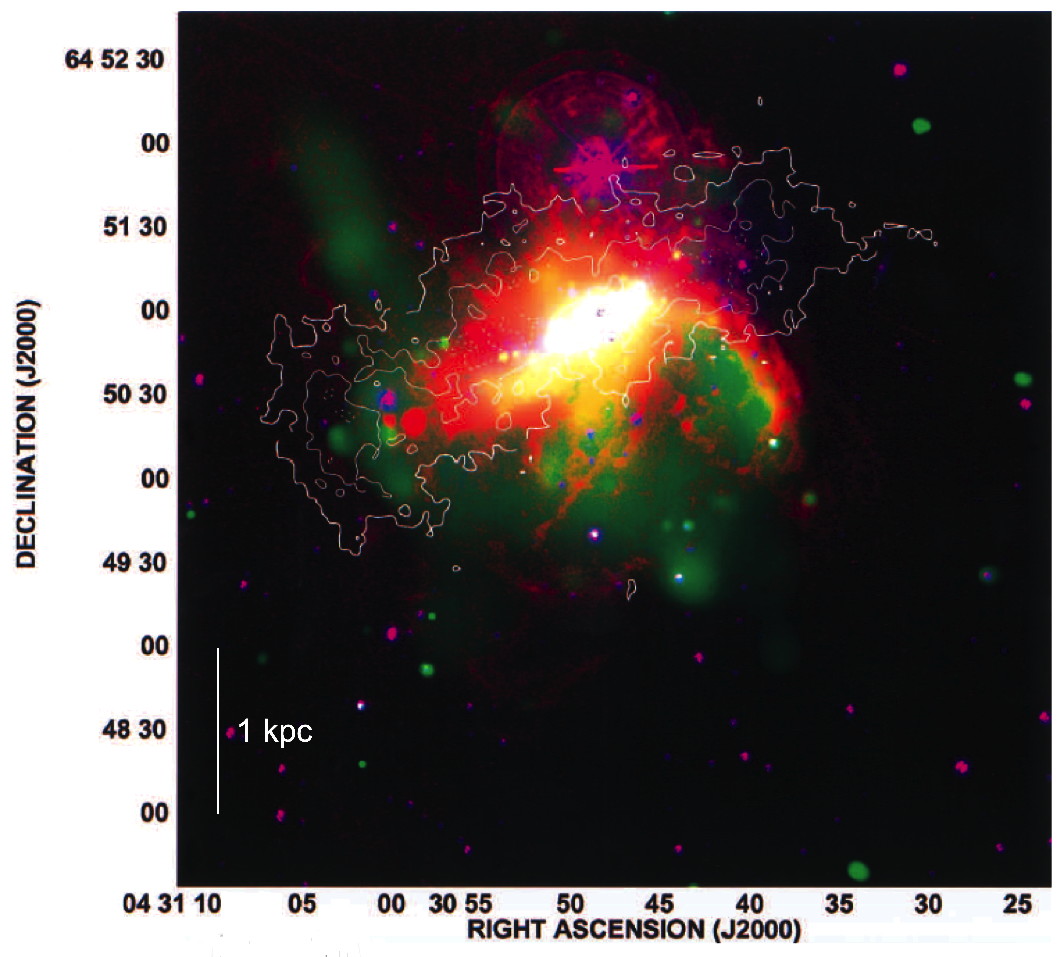}
\end{subfigure}%
\begin{subfigure}{.5\textwidth}
    \centering
    \includegraphics[width=2.3in]{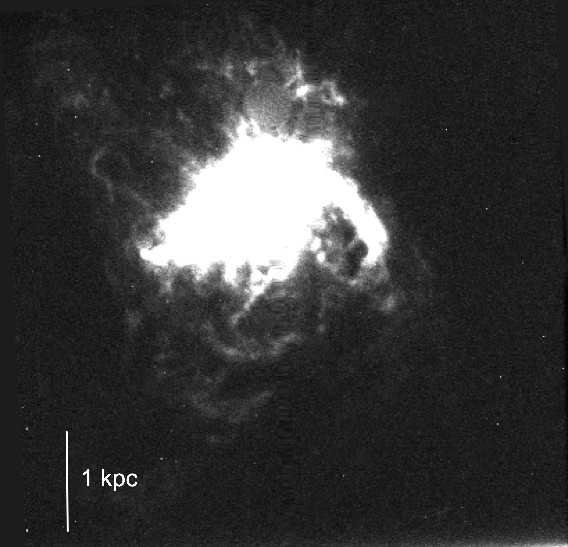}
\end{subfigure}

\caption{
\textit{Left:} 
False color image of NGC 1569 with X-ray (green),
\ha\ (red), 6450 \AA\ continuum (blue), and \HI\ (dashed white line) from
\citet{martin02} with permission of the lead author,
reproduced by permission of the AAS.
\textit{Right:}
\ha\ emission image of NGC 1569 that shows the
large-scale filaments.
}
\label{fig-n1569}
\end{figure}

Intense stellar feedback can result in very large holes that break out of the disk and sometimes
result in outflows into the CGM
\citep{Marasco:23}.
NGC 1569 is a good example of a nearby (2 Mpc) dwarf irregular with massive star clusters \citep{hunter00}, energetic feedback, a kiloparsec-scale hole, and outflow in a hot wind. 
Large-scale (up to several kpc) \ha\ filaments 
extend out of the plane of the galaxy \citep{Hunter:93a,Hunter:93b}.
(See {\bf Figure \ref{fig-n1569}}.)
Spectral analysis by \citet{HandG:97} shows that the large-scale filaments are photoionized, probably by photons escaping \HII\
regions located up to a kpc away. The implied escape fraction is of order 25\%. 
\cite{heckman95} observed that about half of the X-ray flux from this galaxy originates in a $\sim2$ kpc halo, with X-ray spurs along the galaxy minor axis adjacent to the long H$\alpha$ filament. The outflow speed reaches 200 km s$^{-1}$ \citep[see also][]{westmoquette08}, and the outflow age is comparable to the starburst age of $\sim10^7$ years. \cite{heckman95} and \cite{martin98} suggest that the expanding material could eventually blow out into the halo. Spectroscopic observations by \cite{martin02} found that the X-ray wind is metal-enriched, having a solar abundance of $\alpha$ elements, which requires nearly all of the oxygen made in the starburst to go into the wind, whereas the disk \HII\ regions are 0.2$Z$\solar. 
The dust-to-gas ratio (DGR) is higher on the periphery of the \HI\ hole \citep{lianou14}. The magnetic pressure in NGC 1569 is approximately in equilibrium with the other ISM components \citep{kepley10}. 

NGC 1705 \citep{hensler98}, I Zw 18 \citep{martin96,Hunt:05}, NGC 5253 \citep{martin95}, IC 2574 \citep{walter98}, NGC 4214 \citep{hartwell04},
and IC 10 \citep{Heesen:15}
are galaxies similar to NGC 1569 with massive star-forming regions and large bubbles of soft X-ray emission. In some cases, the energy and momentum in the bubble is enough to push the hot gas and local ISM out of the galaxy \citep{marlowe95}. This ``blow-out'' can remove a significant fraction of the metals produced in the starburst \citep{dekel86}, as shown by numerical simulations 
\citep[][for recent work see Emerick et al.\ 2019 and references therein]{deyoung94}
and suggested by observations of NGC 1569 \citep{martin02}, NGC 625 \citep{cannon05b}, I Zw 18 \citep{martin96,Hunt:05}, and Holmberg I \citep{ott01}. Outflows have also been observed in the LMC \citep{ciampa21} and SMC \citep{diteodoro19}. \cite{mcquinn19} observed winds in 6 star-bursting dwarf galaxies using deep H$\alpha$ observations, and \cite{grimes05} observed X-rays from winds in 7 dwarfs. 

Outflows into the CGM, such as seen so dramatically in NGC 1569
and NGC 1705, allow Lyman continuum photons to escape from the cavities \citep{fujita03} and may carry the magnetic field out with it too \citep{chyzy16}. In 43 low-mass galaxies ($10^8M\solar<M_{\rm stars}<10^{10}M\solar$), \cite{bordoloi14} observed halo carbon masses exceeding the carbon masses in the ISM and stars,
implying that considerable quantities of heavy elements are expelled into the CGM in these events.  A metal-rich halo in IC 1613 is also shown by \cite{zheng20}. 
Cosmological simulations that include the multiphase ISM,
star formation, and stellar feedback find large inflow
and outflow rates in the CGM of dwarfs, with a net inflow
of gas from the IGM \citep{li21}. 
The inflow rates are comparable to the SFRs in the disks (see Section \ref{sect-accretion}), suggesting that the two are coupled.
Significant blow-out can also change the gravitational potential in the disk
as gas outflows redistribute mass, causing stars and dark matter to scatter outward 
\citep{elbadry16} and converting a cuspy dark matter center to a core \citep{governato12}.

\subsubsection{Thickness} \label{sect-atomic-thickness}

Dwarf irregular galaxies are generally considered to be disk galaxies, but their disks are thicker relative to their sizes than in spirals
by a factor of 2 to 5 \citep{Brinks:02,Patra:20}.
This is true for stars \citep{Hodge:66,Johnson:17}, gas and dust 
\citep{Walter:99,Dalcanton:04,Roychowdhury:10,EH:15}, and particularly for dwarfs supported more from turbulence than rotation 
\citep{Kaufmann:07,Wheeler:17}.
Less massive dwarfs are apparently puffier than more massive dwarfs \citep{Johnson:17}. 
In a study of \HI\ power spectra, \citet{Zhang:12a} found that LITTLE THINGS dwarfs with absolute magnitudes
$M_V$ brighter than $-14.5$ are relatively thinner than lower mass dwarfs.

The \HI\ in dIrrs also flares or warps more than in spirals \citep{Kepley:07,Patra:14,Szotkowski:19,Patra:20}.
For example, \citet{Banerjee:11} determined \HI\ scale heights for four dIrrs in the THINGS sample. One of the dwarfs
has a scale height of about 450 pc throughout the disk, while the other three flare beyond 3--4 disk scale lengths to as high
as 1 kpc scale height. Flaring is to be expected in most galaxies because the gas velocity dispersion remains approximately constant at large radii while the disk surface mass density declines exponentially.
Deep \HI\ and \ha\ mapping of an edge-on dwarf by \citet{Kamphuis:11} found both neutral atomic and
ionized gas well above the plane of the galaxy: \ha\ gas extends to 655 pc and \HI\ to 1.8 kpc.
In a study of 23 LITTLE THINGS dwarfs, \citet{Patra:20} found that scale heights rise
exponentially with radius from a few hundred parsecs around the centers to a few kpc at the edges.
These thicknesses yield a median axial ratio that is three times that of the Milky Way and
allow stellar radiation to transfer more effectively through the disks of dIrrs than in denser spirals.

Power spectra of the LMC \HI\ reveal a slope transition at $\sim100$ pc scale that suggests
an \HI\ disk of this thickness, with two-dimensional turbulence on larger scales 
having a shallow slope and three-dimensional turbulence on smaller scales having a steeper slope \citep{Elmegreen:01}. The steep part goes to larger scales in the outer regions, suggesting a flare
\citep{Szotkowski:19}.

There are consequences of a thick gas disk on the ability of the dIrrs to form stars.
Relatively thick disks diminish radial instabilities by diluting the radial component of the gravitational force per unit surface density \citep{vandervoort70}. Three-dimensional processes should be relatively more important \citep{EH:15,bacchini20}.
For example, \citet{Elmegreen:11} note that, for a given column density, a thick disk inhibits the instabilities
that can lead to cloud formation, and \citet{EH:15} developed a star formation law based on 3D gaseous gravitational processes
and molecule formation that is more appropriate to dIrrs.

\subsection{Turbulence} \label{sect-atomic-turbulence}

\HI\ linewidths are the result of thermal and bulk motions from turbulence and massive stellar feedback.
Characterization of the \HI\ motions have used third and fourth-order moments, i.e., skewness and kurtosis of the line profiles. \citet{Burkhart:10} compared observations of the SMC to simulations and determined that 90\%
of the \HI\ motions are sub- or transonic. The exception is for supersonic motions around the edges of the bar.
\citet{Maier:17} applied this technique to the LITTLE THINGS galaxies and found that  
turbulent speeds were close to sonic.
In a study of 6 dwarfs using \HI\ power spectra, \citet{Dutta:09} found that several 
showed 2D turbulence on scales larger than the scale-height 
and others showed 3D turbulence.

What drives turbulence in galaxies? Given the obvious energetic feedback from massive stars, turbulence might be expected to come from this feedback energy. Inversely, we might expect turbulence to
compress the gas and initiate new star formation.
The hierarchical clustering in both space and time of young stars, associations, and clusters in the LMC and nearby spiral galaxies
suggests that turbulence helps make star-forming clouds
\citep{efremov98,Grasha:17,Miller:22}.
The velocity dispersion in the ionized gas seems to be the result of thermal motions combined with
young stellar feedback  \citep{Moiseev:15}.

However, spatial correlations between star formation and the kinetic energy density (KED) of the neutral ISM are not obvious.
For example, \citet{Hunter:21a} cross-correlated KED images with SFR density $\Sigma_{\rm SFR}$ images
of LITTLE THINGS dwarfs and found no correspondence. They further compared the excess KED above the radial average with the excess $\Sigma_{\rm SFR}$ 
and again found no correlation. In addition, the excess velocity dispersion in star-forming regions is small.
A similar study of THINGS spirals found the same result \citep{Elmegreen:22}. Their conclusion was that stellar feedback energy
 mostly disrupts molecular clouds without affecting \HI\ turbulence on large scales.
Similarly, \citet{Stilp:13b} compared \HI\ KED measures to time-resolved star formation histories and concluded
that the KED correlates only with star formation from 30--40 Myrs ago, the lifetime of the lowest mass supernova 
progenitor. 
On the other hand, \citet{HunterL:22} found in four nearby dIrrs that on scales of 400 pc, \HI\ turbulence correlates with star formation from 100--200 Myrs ago.

There are other candidates to drive turbulence in the ISM of galaxies, this being an issue for all disk galaxies, not just dwarfs.
\citet{Klessen:10} suggest from simulations that radial accretion is able to drive turbulent motions in spiral galaxies 
if the accretion rate is comparable to the SFR, especially in the outer disk. However, in dwarfs the accretion
rate would have to be much larger than the SFR, so other sources of turbulence are needed.
\citet{Stilp:13a} examined various potential drivers, such as star formation, gravitational instabilities, 
magneto-rotational instabilities, and accretion and concluded that none is capable of driving turbulence in low $\Sigma_{\rm SFR}$ regions, such as dIrrs.

\subsection{Cold HI} \label{sect-coldhi}

The ISM of dIrrs, like that of spirals, is multiphase
with a cold neutral medium 
(CNM, T$\sim$40--200 K in the Milky Way) embedded in a warm neutral medium (WNM, T$\sim$5000--8000 K)
\citep[e.g.,][]{Andersen:00}.
\citet{bialy19}
show, however, that the characteristics
of the multiphase medium depend on the metallicity. 
At very low metallicities of $10^{-3}$--$10^{-2}$$Z$\solar
the cooling rate decreases as metallicity decreases
while the heating rate is nearly independent of the metallicity,
resulting in a need for higher pressures and
densities to form a CNM phase.
At extremely low metallicities of 
$10^{-4}$--$10^{-5}$$Z$\solar\ a multiphase 
medium cannot exist.
Interestingly, at moderate metallicities $>0.1Z$\solar,
the CNM may be colder.

Cold \HI\ may be more important to a galaxy than the total \HI\ since 
the cold \HI\ is more directly connected to star formation \citep{Ian:12}.
Even the tiny dIrr Leo T has a cool \HI\ component in its ISM \citep{Ryan-Weber:08}.
However, observationally there are differences between the cold \HI\ contents in dIrrs and spirals.
In a study of the cold \HI\ in the SMC through absorption spectra, 
\citet{Dickey:00} found that it constitutes only 15\% of the total \HI, 
which is about half that in the solar neighborhood. However, much of this gas is in
dense, $\sim$10--40 K, clouds that would be molecular if they were in the Milky Way. 
\citet{Dempsey:22} found a cold gas fraction of 11\% in the SMC.
We return to this point in Section \ref{sect-COdark}.

Several studies have examined cold \HI\ in dIrrs
by deconvolving the \HI\ line profiles into broad ($\sigma\sim10$ \kms; warm) and narrow ($\sigma\sim2-5$ \kms; cold) 
components.
\citet{Young:96,Young:97} and \citet{Young:03} examined the \HI\ in 7 nearby dIrrs and found that 
the cold component was generally concentrated in a few $10^6$ $M$\solar clumps.
One dwarf, LGS 3, which is notable for having no \HII\ regions, lacked a cold phase altogether.
However, no correlation was found between the quantity of cold \HI\ and the SFR,
and not all cold \HI\ is associated with star forming regions \citep[see also][]{Begum:06}.
\cite{park22} fit cool and warm, bulk (i.e., following the rotation curve) and non-bulk \HI\ to the line profiles in NGC 6822 and found fractions of only 3.8\% and 0.8\% for the cool-bulk and cool-non-bulk components, the rest being warm \HI. 
Using \HI\ power spectra applied to channel maps of LITTLE THINGS dIrrs, \citet{Zhang:12a} found that 
most of the 
cool \HI\ is in the inner disks, although the thermal dispersion of the coolest \HI\ component is smaller than 1.8 \kms, the channel resolution of the data,
making this gas hard to detect (see Section \ref{sect-COdark}).
Cold \HI\ has even been found in the Magellanic Bridge from absorption studies \citep{Kobulnicky:99}.

\subsection{Relationship of \HI\ to Star Formation and the Stellar Disk} \label{sect-atomic-SFandstars}

In dIrrs the azimuthally-averaged surface density of star formation follows the drop with radius of the stellar surface density
better than the gas surface density, even though the gas is the material for star formation \citep{hunter04,Leroy:08}.
Nevertheless, the general relationship between atomic \HI\ and star formation is captured in the Kennicutt-Schmidt relationship
$\Sigma_{\rm SFR}\propto(\Sigma_{\rm gas})^n$ with $n\sim1.3\pm0.3$ 
\citep{Schmidt:59,Kennicutt:89}
where surface density is measured globally or in kpc-size regions.
This relationship was established for spiral galaxies, but \citet{Bigiel:10} extended it to
the outer parts of spirals and dwarfs, regimes of lower SFR and lower gas densities.
They found that the relationship in these extreme regimes is steeper ($n\sim1.7$) than in the inner parts of spirals
and the star formation efficiency (the rate per unit gas) is lower.
On the other hand, a molecular version of the Kennicutt-Schmidt Law shows that $\Sigma_{\rm SFR}\propto\Sigma_{H_2}$
in both spirals and dwarfs \citep{schruba11}.

How far out in radius and how low in gas density can the SFR go?
\citet{Hunter:16} identified the most remote young stellar regions, observable as FUV knots, in 37 LITTLE THINGS galaxies. 
They found regions with ages under 20 Myrs out to 1--8 disk scale lengths, where the minimum \HI\ surface density
was about 1 \unithi.
\citet{Taylor:05} suggested that galaxies with masses $>10^6$ $M$\solar in gas will
form stars until the radiation field heats and stabilizes it
\citep[but see J0139$+$4328 with $8\times10^7$ $M$\solar of gas and no observed stellar component,][]{Xu:23}.
Thus, a tiny galaxy like Leo T, with multiple generations of stars and a gas mass of $4\times10^5$ $M$\solar\ \citep{Ryan-Weber:08}, 
must represent nearly the minimum galaxy for self-regulated star formation.

The Kennicutt-Schmidt relation shows the quantitative correlation between $\Sigma_{\rm SFR}$ and $\Sigma_{\rm gas}$ but it
does not reveal how star-forming clouds form.
The problem with star formation in dIrrs is that the gas densities are low, even in the central regions, and
there are no spiral density waves to pile gas up. 
What mechanisms could make density enhancements in order to form clouds?
An important way to make clouds in spiral galaxies is through gravitational instabilities, which
are quantified in terms of the ratio of a 
critical gas density to the actual gas density, $Q$ \citep{Toomre:64}.
$Q$ is a function of the epicyclic frequency, and so a function of the rotation curve. 
Gas densities above the critical gas density on large scales are unstable and will break up into clouds; gas below the
critical gas density is stable. 
\citet{Hunter:98} found that the \HI\ gas is stable almost everywhere 
in dIrrs. 
In NGC 2366, for example, the star-forming regions are associated with \HI\ peaks that are close
to the critical density even though the average surrounding gas density is below it \citep{Hunter:01}. 
However, unseen molecular gas, i.e., gas that is not clearly associated with CO emission (see Section \ref{sect-COdark}), could make the ISM more unstable \citep{Hunter:19}.

\begin{figure}[t!]
\begin{subfigure}{.5\textwidth}
    \centering
    \includegraphics[width=2.25in]{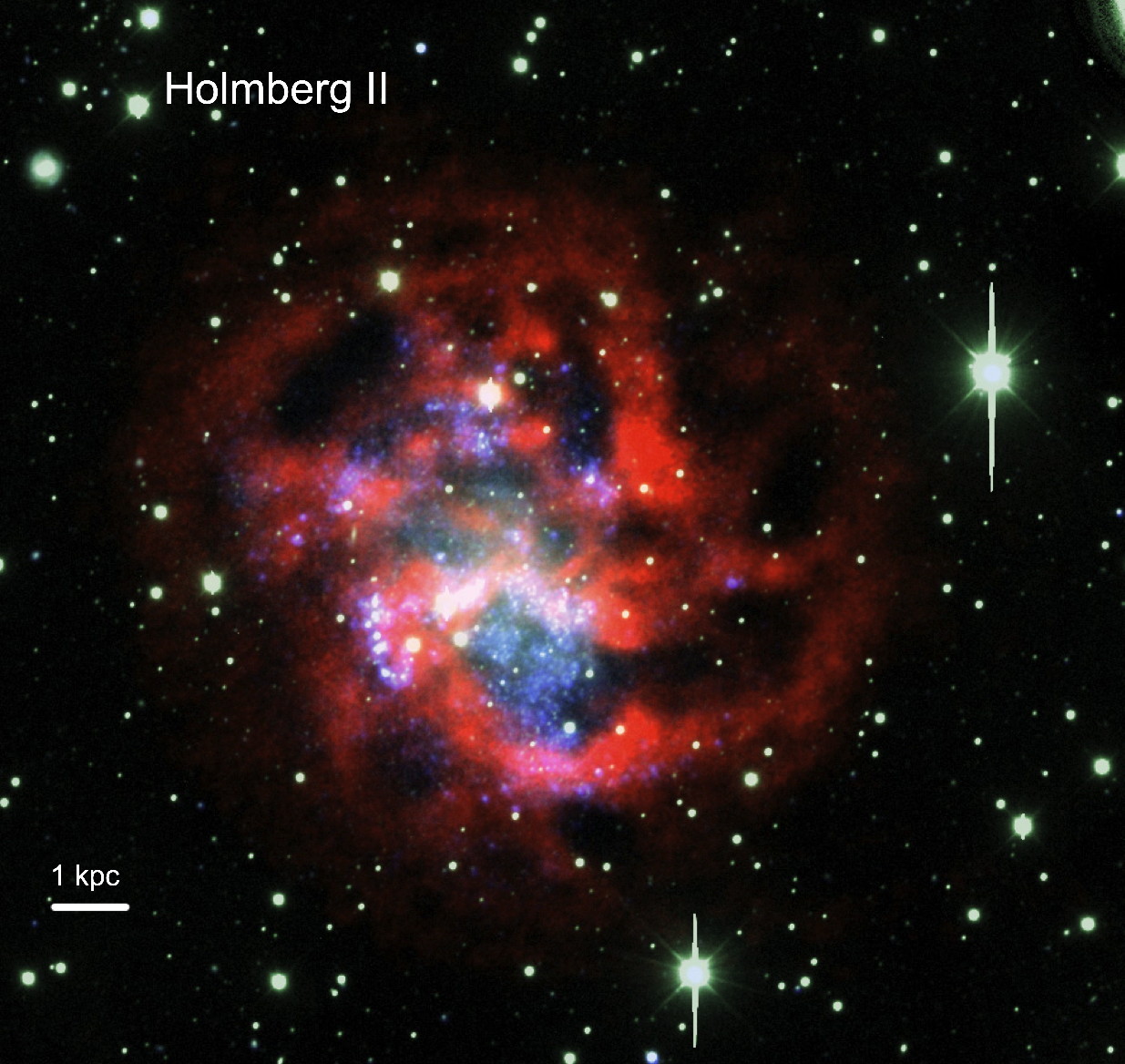}
\end{subfigure}%
\begin{subfigure}{.5\textwidth}
    \centering
    \includegraphics[width=2.25in]{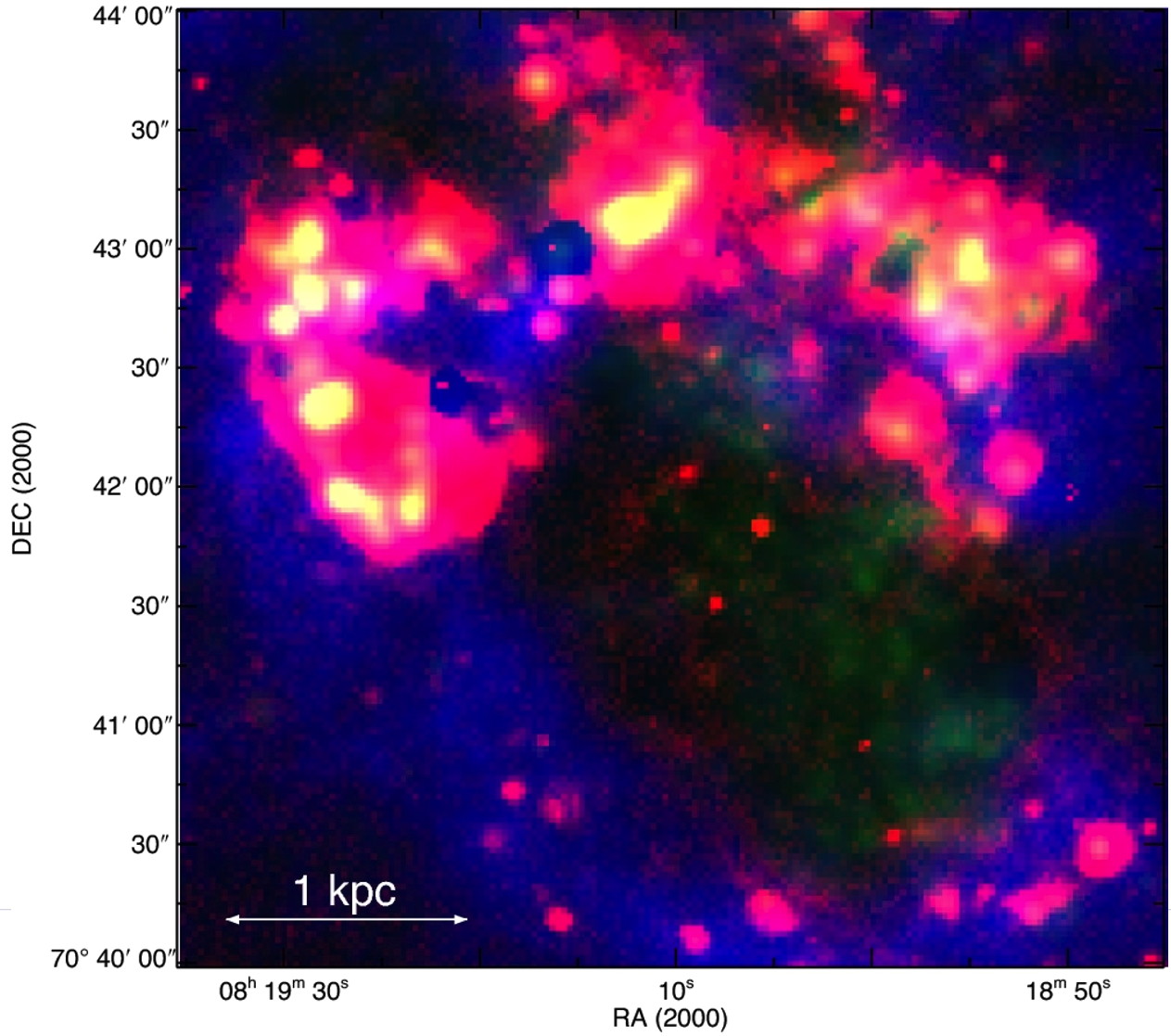}
\end{subfigure}
\caption{
\textit{Left:} 
False-color image of Holmberg II. 
Blue is far-ultraviolet, green is $V$-band, and red is atomic hydrogen gas.
We thank Lauren Hill for producing this image.
\textit{Right:}
Giant gas shell with peripheral star formation in Holmberg II, with red, green and blue corresponding to H$\alpha$, FUV (GALEX) and \HI\ 21-cm (LITTLE THINGS) emission, respectively. 
This shell is centered on the prominent hole just south of the center of the galaxy.
From \citet{egorov17} with permission from the lead author and permission of OUP. 
}
\label{fig-egorov17}
\end{figure}

There are other processes for 
producing gas density enhancements that could be important in dwarfs:
stellar feedback, turbulence, and external gas accretion. 
As discussed in Section \ref{sect-atomic-holes}, 
feedback from massive stars can pile gas into shells and trigger the formation of a next generation of stars (see {\bf Figure \ref{fig-egorov17}}).
This process does take place in dIrrs, but 
only a small percentage ($<15$\%) of the star formation in each galaxy is located in obvious shells \citep{Bolds:20}.
Turbulence is another way to make density enhancements that become self-gravitating, but
there is no direct evidence for particular star-forming events triggered in this way (see Section \ref{sect-atomic-turbulence}). On the other hand, cloud-cloud collisions may be considered a form of turbulence, and there are many examples suggesting cloud collisions trigger star formation \citep[see review in][]{fukui21}.
It has also been proposed  that accretion of gas from the CGM or IGM furthers star formation 
\citep[see, for example, simulations by][]{li21},
but, again, there is no
observational evidence for wide-spread accretion of this sort onto dwarfs (or spirals; however, see the example of tadpole galaxies in Section \ref{sect-accretion}).
Star formation is a local process that depends on global conditions \citep{Hunter:08} but aside from the formation of giant shells, we have yet to 
find observational evidence for what processes are most important in initiating the local conditions that are
the first step to making star-forming clouds. Perhaps cloud formation is from a combination of processes that are difficult to recognize at the initial stages.

Because stars form out of the gas, we expect that features in the
stellar disk would be accompanied by changes in the gas disk.
One feature of stellar disks of both spirals and dwarfs is that the stellar surface brightness profiles frequently show
a ``break'',
where the exponential fall-off with radius abruptly changes slope \citep[for example, in dwarfs see][]{Herrmann:13}. 
In most disks the bend is downward
(Type II) but in some it is upward (Type III). Few disks show no break (Type I). 
For dIrr galaxies these surface brightness breaks are also present in stellar mass density profiles.
So what happens at the break? 
There is no correlation between the \HI\ surface density or features of
the gas rotation curves and the location of the break \citep{Hunter:21b}.
Currently, we do not understand the connection between the breaks in stellar exponential disks and the star formation and gas 
characteristics.


\section{STAR-FORMING REGIONS AND IONIZED GAS} \label{sect-SFingregions}

Most dIrr galaxies have \HII\ regions where star formation takes place and dense clouds are ionized
by young massive stars. The \HII\ region luminosity function is a power law for dIrrs as it is for spirals, but
the power law is shallower in dIrrs (slope$\sim -1.5\pm0.1$ for those with turnovers or $\sim -1\pm0.1$ for those without) than in spirals ($\sim -1.9$)
\citep{Youngblood:99,thilker02,oey03}. 
Also, the upper luminosity cutoffs are lower in dIrrs, which do not often have supergiant \HII\ regions.
However, nearby dIrrs do frequently have complexes of smaller \HII\ regions
compared to nearby spiral galaxies.
In fact most of the H$\alpha$ luminosity from \HII\ regions comes from regions that are smaller than 10 times the Orion nebula.
Fourier transform power spectra of dIrr H$\alpha$ emission, taken along the major axes, have revealed a
universal slope inside \HII\ regions that is consistent with Kolmogorov turbulence \citep{Willett:05}.
\citet{cormier15} also found that \HII\ regions in dwarfs have harder radiation fields than
in spirals due to the impact of lower metallicity on stellar emission, 
specifically main sequence temperatures are
hotter due to reduced line blanketing and blocking 
in stellar atmospheres
\citep{madden06}.


In addition, dIrrs, like spirals, contain diffuse ionized gas (DIG) outside of \HII\ regions.
Analysis of the DIG in 14 dIrrs by \citet{Martin:97} showed that most of the DIG is photoionized by
massive stars in star forming regions and is a result of the porosity of the ISM in these galaxies
\citep[see also][]{Polles:19,Hidalgo-Gamez:07}.
\citet{Choi:20} showed that the fraction of ionizing photons escaping the galaxy is about 25\% for NGC 4214.
Multiphase modeling of the ISM of dwarfs by \citet{Ramambason:22} \citep[see also][]{Cormier:19} showed that the structure
and porosity varies with metallicity. The escape fraction of ionizing photons goes up as the metallicity
goes down, with escape fractions reaching as high as 60\%. 
This suggests that dIrrs and similar systems could provide ionizing photons to the IGM
to be a source of reionization in the early universe \citep{enders23}.

In some dIrrs there are also embedded star-forming regions that radio continuum or IR observations
have revealed \citep[for example][]{Hindson:18}.
NGC 5253 contains a large \HII\ region and massive star cluster with thousands of O-type stars 
that are detected in 7 mm continuum and near-IR \citep{Turner:04}.
Mid-IR observations of NGC 1569 have revealed an embedded compact source that is optically thick in 
the radio and may contain nearly a thousand O-type stars \citep{Tokura:06}.
Centimeter radio observations of NGC 4449 have revealed numerous thermal radio sources with ages
less than 5 Myr and stellar masses of order $10^4$ $M$\solar \citep{Reines:08,calzetti23}.
These are all large star-forming regions found in small galaxies with high SFRs.

Coronal line emission has been found in dwarfs as part of the Sloan Digital Sky Survey (SDSS), with the
dwarfs preferentially having emission lines with the highest ionization potentials \citep{Reefe:22}.
These emission lines could come from central black holes with hotter accretion disks than those around 
supermassive black holes in spiral galaxies.
\citet{Bohn:21} identified 5 galaxies at the massive end of the dwarf range with coronal line emission 
that is likely due to AGNs accompanied by outflows,
and \citet{hatano23} have found the signature of a dust torus
around a massive black hole in the dwarf SBS0335-052E.
The relative population of AGNs in dwarfs and the effect of AGN activity on 
their ISM are not known.


\section{MOLECULAR GAS} \label{sect-molecular}
\subsection{Observations} \label{sect-molecular-obs}

In the Milky Way we see that star formation takes place in clouds of molecular gas. For that reason, molecular gas is crucial to the evolution of galaxies. For the reasons outlined below, observations of molecular gas in dIrrs have been very difficult.

\subsubsection{Early Discoveries} \label{sect-molecular-early}

CO is weak in low metallicity galaxies. It took over a decade longer to detect CO in the SMC \citep{israel86a} than in the LMC \citep{huggins75} with metallicities of $0.2Z_\odot$ and $0.5 Z_\odot$, respectively. Early low-resolution surveys \citep{cohen88,rubio91} and high resolution maps \citep{israel93,rubio93a,rubio93b} of the Magellanic Clouds eventually demonstrated the influence of metallicity, dust, and UV radiation on CO emission. \cite{lequeux94} concluded that the CO parts of dense cloud complexes in the SMC shrink to the inner cores while \HI\ and possibly H$_2$ surround them. This is still the primary model at low metallicity, discussed now in terms of elevated ratios of H$_2$ column density to integrated CO line strength, $X_{\rm CO}$, and molecular surface density to integrated CO line strength, $\alpha_{\rm CO}$, or, equivalently, molecular mass to CO luminosity. 

The third dIrr detection, NGC 1569, was made with the 14-m telescope at the Five College Radio Astronomy Observatory by \cite{young84}. This galaxy was soon followed by NGC 3738, NGC 4214, and NGC 4449 \citep{tacconi85},  and six more, including 3 BCDs, in a survey of 15 dIrrs by \cite{tacconi87}. At the same time, CO in IC10 was discovered by \cite{henkel86} with the 7-m telescope at AT\&T Bell Laboratories, and mapped in 5 points by \cite{ohta88} with the 45-m telescope of the Nobeyama Radio Observatory (NRO). The first detection in a galaxy smaller than the LMC (other than the SMC), NGC 3077, was made with the 30-m IRAM telescope by \cite{becker89}. 

\begin{marginnote}[]
\entry{$X_{\rm CO}$}{Ratio of H$_2$ column density to integrated CO line strength.}
\entry{$\alpha_{\rm CO}$}{Ratio of average molecular surface density to integrated CO line strength or molecular mass to CO luminosity for unresolved clouds.}
\end{marginnote}

Larger early surveys included NGC 6822 by \cite{wilson92}, who used the NRAO 12-m telescope, 13 BCDs and 2 dIrrs by \cite{sage92} with IRAM, NGC 6822 and NGC 4214 by \cite{ohta93} with NRO, He 2-10 by \cite{baas94} with the 15-m SEST telescope and by \cite{kobulnicky95} with the first CO interferometry using Owens Valley Radio Observatory.  A few more dIrrs by \cite{israel95} with SEST and NRAO telescopes, in addition to \cii\ observations of IC 10 by \cite{madden97} with NASA’s Kuiper Airborne Observatory, led these authors to conclude that significant H$_2$ in dIrrs was unseen. 

In the two decades since these early discoveries, there have been many observations of molecules of various types in dIrrs. In the first large survey, \cite{leroy05} searched for CO(1-0) in 121 galaxies of average dynamical log mass $\sim9.6$ (in $M_\odot$) and detected 28; the lowest-mass detections were NGC 5338 (Hubble type SB0) and NGC 3913 (Sc).
\cite{schruba12} observed 16 local dwarfs in CO(2-1) and detected  5 by stacking the spectra. \cite{Cormier:14} surveyed 6 dwarfs in several CO transitions and detected 5, all with metallicities above that of the SMC; their one galaxy with SMC-type metallicity, NGC 4861, was undetected.

The most recent SMC surveys are in \cite{saldano23} at 9 pc resolution and \cite{ohno23} at 2 pc. Ohno et al.\ found that molecular clouds in the SMC have a mass spectrum similar to that of the Milky Way, with a slope for $dN/dM$ equal to $\sim-1.7$, and they have slightly narrower linewidths for their size, $\sigma_v\sim0.5R^{0.5}$ for radius $R$ in pc and velocity dispersion $\sigma$ in km s$^{-1}$, compared to $\sim0.7R^{0.5}$ in the Milky Way. Salda\~{n}o et al.\ derived a CO to H$_2$ conversion factor $\alpha_{\rm CO}=10.5\pm5\;M_\odot$ (km s$^{-1}$ pc$^2$)$^{-1}$ assuming virialized clouds for the total mass and $28\pm15\;M_\odot$ (km s$^{-1}$ pc$^2$)$^{-1}$ using the dust emission for the mass. These are larger than in the Milky Way by factors of 2.4 and 6.5.

\subsubsection{The Lowest Metallicities} \label{sect-molecular-lowest}

CO has been mapped in only a few galaxies with metallicities comparable to or lower than that of the SMC. The observations are summarized in what follows with an emphasis on unique features for each galaxy. 
Generally, pc-scale observations are necessary to resolve individual CO clouds, which tend to be much smaller than their CO-dark envelopes and generally a few pc in size.
If we consider the large-scale gas concentrations to be composite clouds, then the ratio of this cloud mass to the mass of the CO-luminous part can be in the hundreds for dIrrs -- even larger for the lowest metallicities, suggesting extensive peripheral gas unobserved in \HI, H$_2$, or CO. This peripheral gas is still visible in other forms, such as dust or \cii\ emission (see below), but it is not as prominent in CO as it is in the Milky Way. Nevertheless, the CO parts of low-metallicity clouds are very similar to scaled-down versions of CO clouds elswehere: they have comparable densities and extinctions, and they are close to virialized.  

Shrinking CO cores inside dense cloud complexes can be characterized by variations in the CO(1-0)-to-H$_2$ conversion factor.  Galactic conversion factors are typically taken to be $X_{\rm CO} = 2.0\times10^{20}$ H$_2$ molecules cm$^{-2}$ (K km s$^{-1}$)$^{-1}$ and $\alpha_{\rm CO} = 4.3\;M_\odot$ (K km s$^{-1}$ pc$^2$)$^{-1}$  
where the latter includes a factor of 1.36 for He and heavy elements \citep{Bolatto:13}. 
Note that $\alpha_{\rm CO}$ written this way may be viewed as a total cloud mass per unit CO luminosity, $L^\prime$ in K km s$^{-1}$ pc$^2$, but is equivalently written as a total gas surface density per unit CO line integral with units of $M_\odot$ pc$^{-2}$ (K km s$^{-1})^{-1}$. The first way allows one to consider undetected gas surrounding the CO source, as in a cloud envelope with CO in the core.

Derivations and caveats with the conversion factors are discussed in Section \ref{sect-conversion}. These factors tend to increase with decreasng metallicity, clumpiness, and increasing physical scale, as the CO-emitting gas becomes a smaller fraction of the total. They also increase with specific star formation rate (sSFR), as the excess FUV photons penetrate further into the cloud envelopes, disassociating the molecules more (Sect. \ref{sect-conversion-theory}). These dependencies make it ambiguous to plot a quantity like $\alpha_{\rm CO}$ against just one parameter representing dIrrs, such as metallicity.

{\bf NGC 6822} ($12+\log \textrm{O/H}=8.02$ or $0.21Z_\odot$, distance $D=474$ kpc): \cite{schruba17} mapped CO(2-1) in four 250 pc regions using the Atacama Large Millimeter/submillimeter Array (ALMA)
at 2.0 pc (FWHM) and 0.635 km s$^{-1}$ resolutions. They found $\sim150$ small CO cores with average 2.3 pc radius, 1.1 km s$^{-1}$ linewidth, $2.7\times10^3\;M_\odot$ virial mass, and $125\;M_\odot$ pc$^{-2}$ surface density, concentrated inside $\sim10^6\;M_\odot$ dust$+$gas complexes. The conversion factor for CO emission to total molecular mass was derived from the dust emission corrected for \HI\ to be $\alpha_{\rm CO}=85\pm25\;M_\odot$ (K km s$^{-1}$ pc$^2$)$^{-1}$ in the two fields with compact H$\alpha$ and high CO luminosities, and it was derived to be much larger, $235\pm72$ and $572\pm93\;M_\odot$ (K km s$^{-1}$ pc$^2$)$^{-1}$ in the two fields with extended H$\alpha$ and low CO luminosity. For the 2.3 pc CO cores themselves, $\alpha_{\rm CO}$ was within a factor of 2 of the Milky Way value and the cores looked like normal Milky Way clouds: the H$_2$ densities calculated from the virial theorem, pressure balance and excitation are all about the same, $10^3$ cm$^{-2}$, and the extinctions through the H$_2$ envelope and CO core, 0.5 mag and 2.3 mag, are comparable to Milky Way values. 

{\bf DDO 50} ($12+\log \textrm{O/H}=7.92$ or $0.17Z_\odot$, $D=3.27$ Mpc): \cite{shi16} detected CO(2-1) from two sources in DDO 50 with the IRAM 30m telescope at a resolution of 170 pc.  The emission line in one of them, DDO 50A, was detected at a signal-to-noise ratio of $5.9\sigma$ and 18 km s$^{-1}$ FWHM; in the other, DDO 50B, there were two components with combined $6.1\sigma$ detection and 18 km s$^{-1}$ FWHM. Using the estimated dust mass from FIR emission minus the observed \HI\ mass to determine the total molecular mass, the conversion factors were derived to be  $546^{+1095}_{286}$ and $302^{793}_{202}$ in $M_\odot$ (K km s$^{-1}$ pc$^2$)$^{-1}$.

{\bf NGC 2366} ($12 + \log \textrm{O/H} = 7.89$ or $0.16Z_\odot$; $D=3.4$ Mpc):  
\cite{oey17} used the Northern Extended Millimeter Array (NOEMA) 
at $7.9 ~{\rm pc}\times5.8$ pc resolution to detect 5 CO(2-1) clouds around Mrk71-A, which is a young super-star cluster of mass $1.4\times10^5\;M_\odot$ in NGC 2366.  The average cloud radius is $\sim4.6$ pc and the virial mass is $10^5\;M_\odot$, making the average virial density $\sim5000$ cm$^{-3}$ and extinction 14 mag. The average virial mass is close to the average luminous mass, $1.26\times10^5\;M_\odot$ using an assumed conversion factor of X$_{\rm CO}=50\times10^{20}$ H$_2$ cm$^{-2}$ (K km s$^{-1}$)$^{-1}$. Taking the same ratio to the Milky Way values, this becomes $\alpha_{\rm CO}=107 M_\odot$ (K km s$^{-1}$ pc$^2$)$^{-1}$.

{\bf Kiso 5639} ($12+\log \textrm{O/H}=7.83$ or $0.14Z_\odot$, $D=24.5$ Mpc): Kiso 5639 is a lopsided dwarf starburst of the tadpole morphology ($M_{\rm star}\sim 5\times10^7\;M_\odot$) with a giant star-forming region on one side, 300 pc across and having a SFR of $\sim 0.04\;M_\odot$ yr$^{-1}$. \cite{elmegreen18} mapped it in CO(1-0) with NOEMA at $340 ~{\rm pc}\times440$ pc and 5.1 km s$^{-1}$ resolutions. 
Assuming $\alpha_{\rm CO}=100\;M_\odot$ (K km s$^{-1}$ pc$^2$)$^{-1}$, 
they derived a molecular cloud mass of $2.9\times10^7\;M_\odot$ and $\Sigma_{\rm CO}=120\;M_\odot$ pc$^{-2}$ with two primary components separated by 19 km s$^{-1}$. This internal molecular velocity spread is a large fraction of the total galaxy rotation speed, 35 km s$^{-1}$, suggesting a catastrophic event connected with either the formation or the destruction of the molecular cloud. The molecular cloud itself is not extraordinary aside from its large mass. Using the deconvolved radius and velocity dispersion, the virial parameter is 1.8 (less than 1.1 for each component), and the surface density corresponds to $A_V=0.8$ mag. The pressure in the cloud is also reasonable for a starburst region, $4.8\times10^5k_B$, which is like the pressure in Milky Way giant molecular clouds (GMCs).

\begin{figure}[t!]
\begin{subfigure}{.30\textwidth}
    \centering
    \includegraphics[width=1.5in]{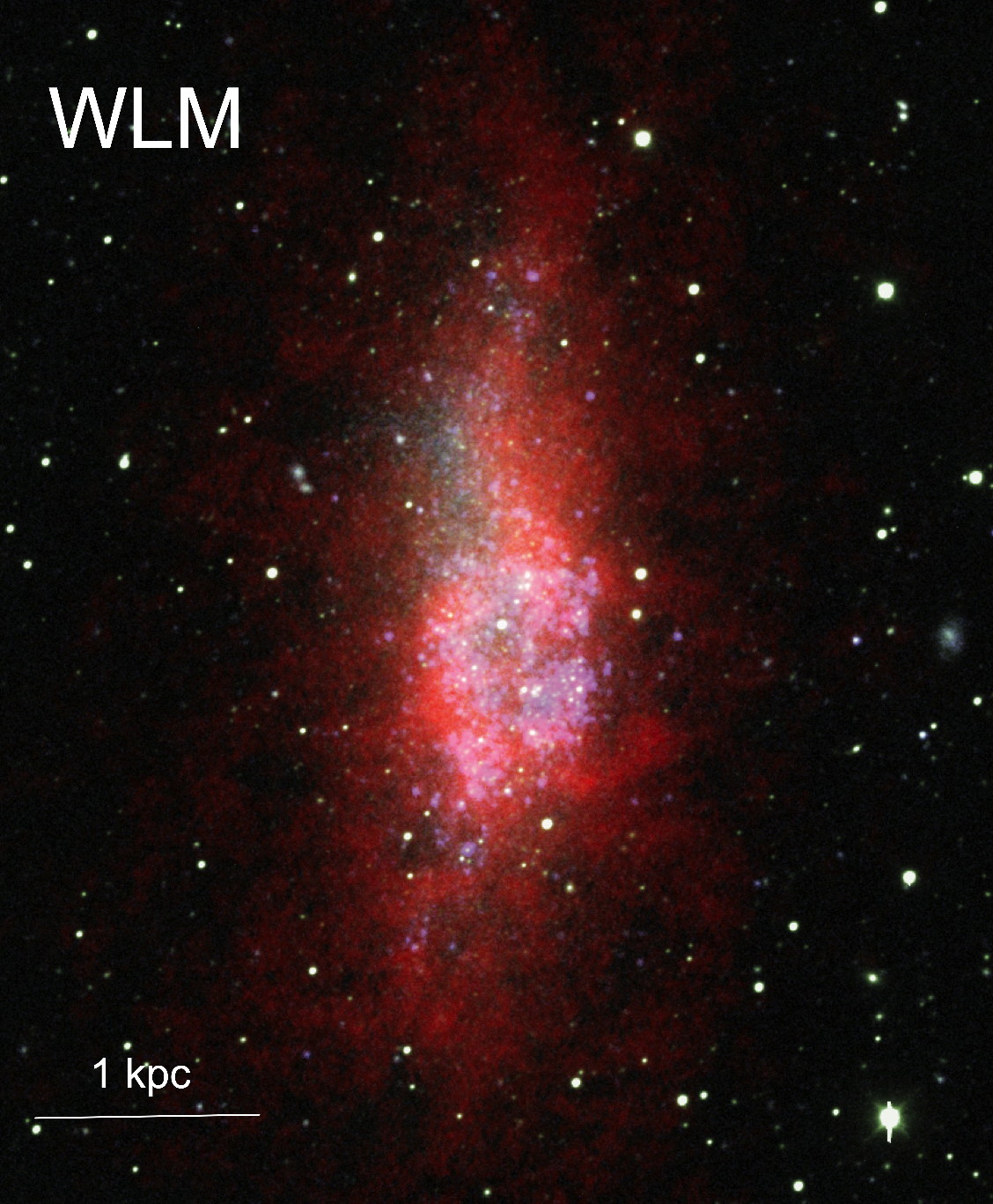}
\end{subfigure}%
\begin{subfigure}{.37\textwidth}
    \centering
    \includegraphics[width=1.8in]{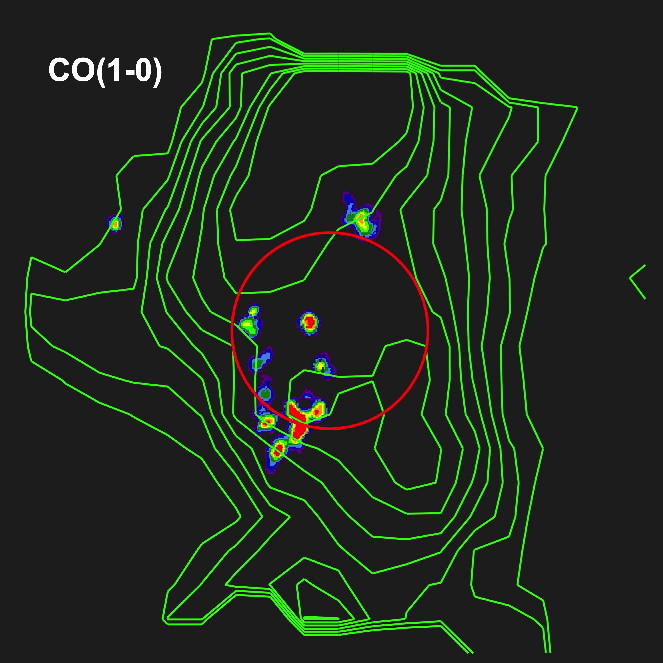}
\end{subfigure}%
\begin{subfigure}{.37\textwidth}
    \centering
    \includegraphics[width=1.8in]{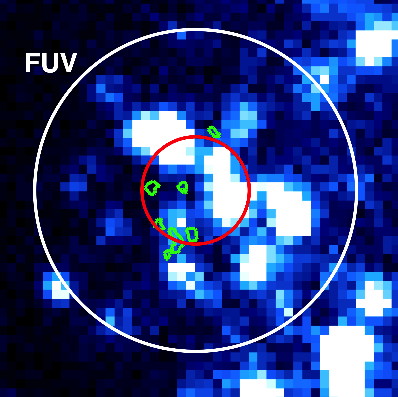}
\end{subfigure}
\caption{
\textit{Left:} False-color image of WLM.
Blue is far-ultraviolet, green is $V$-band, and red is atomic hydrogen gas.
We thank Lauren Hill for producing this image.
\textit{Middle and Right:} Star-forming region in WLM at 0.13$Z\solar$. The red circle has
a diameter of 18\arcsec = 86 pc.
\textit{Middle:} False-color ALMA map of CO (1-0) 
\citep[data from][used with permission from the lead author]{Rubio:15}.
{\it Herschel} [CII]$\lambda$158 micron emission ({\it green contours}) defines the PDR \citep[data from][used with permission from the lead author]{cigan16}.
\textit{Right:} GALEX FUV image. Small green contours outline the CO cores.
}
\label{fig-wlmse}
\end{figure}

{\bf DDO 53} ($12+\log \textrm{O/H}=7.82$ or $0.13Z_\odot$, $D=3.68$ Mpc):
\cite{shi16} detected the CO(2-1) line in DDO 53 at $S/N=7.1\sigma$ with the IRAM 30m at 200 pc resolution. Using the estimated dust mass minus the observed \HI\ mass to determine the total molecular mass, they derived a conversion factor of $\alpha_{\rm CO}=261^{+940}_{-249}\;M_\odot$ (K km s$^{-1}$ pc$^2$)$^{-1}$.

{\bf WLM} ($12+\log \textrm{O/H}=7.8$ or $0.13Z_\odot$, $D=985$ kpc): \cite{Rubio:15} used ALMA to map CO(1-0) in two $\sim300$ pc regions of WLM with $6.2 ~{\rm pc}\times4.3$ pc spatial resolution (HPBW) and 0.5 km s$^{-1}$ velocity resolution (FWHM). They found 10 small clouds with an average radius of 2.5 pc and an average virial mass of $2.8\times10^3\;M_\odot$
(see {\textbf{Figure \ref{fig-wlmse}}}).
The individual clouds satisfied the usual correlations for GMCs in the Milky Way \citep{larson81}, namely linewidth versus size and mass versus radius, although they occupied the low-mass and low-radius ends of these correlations. The CO clouds are also approximately virialized, considering that the H$_2$ density required for virial equilibrium is about equal to the density required for CO excitation (i.e., $\sim10^3$ cm$^{-3}$), and this density multiplied by the square of the linewidth gives an internal pressure about equal to the cloud boundary pressure derived from the weight of the overlying gas as observed in dust emission. The extinctions in the \HI+H$_2$ envelope and the CO cores themselves are also about equal to the threshold extinctions for H$_2$ formation ($A_V=0.3$ mag and 1.5 mag, respectively), considering the conversion of gas surface density to extinction, assuming the DGR  scales with metallicity. Further study of the WLM clouds and their environments is in \cite{archer22},
where they find no obvious environmental characteristics driving the formation of CO cores in WLM.

{\bf Sextans A} ($12+\log(\textrm{O/H})=7.54$ or $0.07Z_\odot$, $D=1.4$ Mpc):
A marginal detection of CO(1-0) at $S/N=3.4\sigma$ is reported in \cite{shi15} in the main star-forming region of Sextans A with the IRAM 30m telescope at 150 pc resolution. Using the derived dust mass minus the observed \HI\ mass, the molecular mass was determined to be $(1.0\pm0.3)\times10^7\;M_\odot$, which implies $\alpha_{\rm CO}=(2.8\pm1.1)\times10^3\;M_\odot$ (K km s$^{-1}$ pc$^2$)$^{-1}$.

{\bf Sextans B or DDO 70} ($12+\log(\textrm{O/H})=7.53$ or $0.07Z_\odot$, $D=1.38$ Mpc):
\cite{shi16} detected CO(2-1) in Sextans B at $S/N=5.5\sigma$ with 74 pc resolution on the IRAM 30m telescope. Using the dust mass minus the \HI\ mass, they derived $\alpha=6949^{+23403}_{-6067}\;M_\odot$ (K km s$^{-1}$ pc$^2$)$^{-1}$. With ALMA, \citet{shi20} also mapped the CO(2-1) emission at 1.4 pc and 0.4 km s$^{-1}$ resolutions. They found 5 clumps with radii of 1.5 to 3 pc and velocity dispersions of 0.6 to 1 km s$^{-1}$ that satisfied the Milky Way size-linewidth relation, and they derived virial masses of $0.7\times10^3$ to $3\times10^3\;M_\odot$. As for other low-metallicity galaxies discussed above, the extinction required to produce the CO is $A_V\sim1$ mag, corresponding to gas mass surface densities of $756\pm467\;M_\odot$ pc$^{-2}$. At these mass column densities, Milky Way CO clumps are $\sim4$ times smaller. 

{\bf I Zw 18} ($12+\log(\textrm{O/H})=7.18$ or $0.03Z_\odot$, $D=18.2$ Mpc):  \cite{zhou21} tentatively detected CO(2-1) in I Zw 18 at $S/N=3\sigma$ using NOEMA with 150 pc resolution. The total CO luminosity was found to be $\sim4\times10^3$ K km s$^{-1}$ pc$^2$, which is not the lowest detected, as it was lower in two of the galaxies mentioned above (Sextans A: $3.7\pm1.0$, DDO 70: $0.204\pm0.037$ in units of $10^3$ K km s$^{-1}$ pc$^2$). I Zw 18 has $\sim100$ times higher SFR for its CO luminosity than these others. Normalized to the sSFR as in \cite{hunt20}, the ratio of the CO luminosity to the SFR is consistent with the others for its metallicity.

Among these nine dIrrs, four have been observed with a spatial resolution of several parsecs (NGC 6822, NGC 2366, WLM, Sex B), three have clear detections on spatial scales of several hundred pc, and two have marginal detections on these larger scales. For the high resolution observations, the CO clouds have sizes of only a few parsecs; in three of them, the average CO cloud mass is several $\times10^3\;M_\odot$ and in the fourth (NGC 2366) the average mass is $\sim10^5\;M_\odot$. Also for the high-resolution observations, the CO cloud densities and extinctions are adequate for molecular excitation and survival, and the cloud sizes and linewidths lie close to the relationship for the Milky Way.

It is not clear from these observations how small clouds in the more quiescent dIrrs make massive stars, which must be present to explain the H$\alpha$ emission and the feedback that produces giant shells. Small clouds apparently also make star clusters up to several hundred solar masses, which are present in these galaxies as well. The much larger molecular mass in Kiso 5639 is presumably what allows such regions to make super star clusters and perhaps globular clusters in the early universe.


\subsubsection{Molecules in Other Dwarf Galaxies} \label{sect-molecular-other}

Dwarf galaxies with metallicities higher than that of the SMC tend to have more CO emission per unit molecular mass and cloud structure approaching that in the Milky Way, where the CO parts of most clouds have relatively small molecular envelopes and H$_2$ is present without CO in the denser diffuse regions \citep{lee12}.

In the LMC, the Columbia \citep{cohen88}, NANTEN \citep{fukui08}, and MAGMA \citep{hughes10} CO(1-0) surveys at resolutions of 130 pc, 40 pc, and 8 pc, cataloged 40, 272, and 175 GMCs, respectively, finding $X_{\rm CO}$ factors from 3 to 6 times higher than in the Milky Way using the virial theorem. The clouds have slightly narrower linewidths for their sizes than in the Milky Way but the mass spectrum is the normal power law \citep[see also][]{brunetti19}.  \cite{herrera13} measured $X_{\rm CO}$ more precisely by converting the dust emission into a gas mass in the N11 star-forming region and comparing that to CO luminosity and cloud virial mass. \cite{chevance20} mapped the FIR lines toward 30 Doradus and applied a photodissociation region (PDR) model to determine the total molecular gas mass. They found that 75\% of the molecular gas mass is not traced by CO. 

Individual clouds in the LMC are not much different from Milky Way clouds. \cite{naslim18} used ALMA to observe clouds in the N55 region at 0.67 pc resolution and found them to be in virial equilibrium with an $X_{\rm CO}$ factor about twice the Milky Way value.  \cite{wong19} showed that the CO linewidths in 6 LMC clouds increase with 8$\mu$m radiation intensity and surface density for a given size.  Dense gas emission from HCO$^+$ and HCN correlates with star formation as in other galaxies \citep{galametz20,nayana20}. In the 50 pc region around the 30 Doradus star-forming complex, \cite{wong22} mapped CO(2-1) and $^{13}$CO(2-1) with ALMA at 0.4 pc resolution and found virialized clumps in filamentary structures \citep[see also][]{indepetouw13,anderson14}. In the southern part of the molecular ridge below 30 Doradus, \cite{finn21} observed quiescent CO clumps and found a correlation between cloud density and the presence of young stars. 

In other local dwarfs with metallicities of a few tenths solar or higher, especially in dwarf starbursts and mergers, CO is a good diagnostic for 
interstellar pressure and star formation feedback. Dwarf mergers (Section \ref{sect-mergers}) can have bright CO emission with higher localized star formation activity than normal spirals. For example, \cite{kepley16} mapped CO(3-2) and CO(2-1) in the merger II Zw 40 ($0.25Z_\odot$, 10 Mpc) at 24 pc resolution and found several clouds with one containing significant star formation. The clouds have higher linewidths for their sizes than Milky Way clouds, corresponding to higher gas surface densities and pressures from the merger.  \cite{consiglio16} also mapped CO(3-2) and CO(1-0) in II Zw 40 at 20 pc resolution, and they mapped 870$\mu$m dust emission. They found that the DGR varies with position by a factor of $\sim4$, with higher values closer to the starburst as a result of dust production by large numbers of massive stars. 
\cite{gao22} studied CO(1-0) and \ci\ in the merger starburst Haro 11 ($\sim0.27Z_\odot$, 87.2 Mpc) at 460 pc resolution, and measured elevated sSFRs and star formation efficiencies in the active clumps, with values comparable to those in high redshift galaxies and local ULIRGS. 
In the starburst Haro 2 ($0.3Z_\odot$, 21 Mpc), which may not be a recent merger, \cite{beck20} observed CO(2-1) at 200 pc resolution and found that half the molecules are in two large clumps with embedded star formation and the other half are in an expanding X-ray bubble. 

The large impact that gas accretion can have on dwarf galaxies (Section\ \ref{sect-accretion}) is also evident in CO observations. 
\cite{beck18} observed He 2-10 ($0.8Z_\odot$, 8.7 Mpc) in CO(3-2) with ALMA at 12 pc resolution, finding filaments that seem to be feeding the starburst and another region 25 pc in size with an outflow energy in excess of $10^{53}$ erg. \cite{imara19} observed the same galaxy in CO(1-0) and identified 119 virialized clouds with deconvolved sizes of $\sim26$ pc and surface densities of $\sim180$ $M$\solar pc$^{-2}$, resembling Milky Way clouds but with 50\% larger linewidths for their size, indicating larger pressures. \cite{miura18} observed the nearest starburst dwarf NGC 5253 ($Z\sim0.3Z_\odot$, 3.15 Mpc) in CO(2-1) at 3 pc resolution with ALMA. NGC 5253 is also accreting molecular gas \citep{miura15}. They found 118 molecular clouds with an average radius of 4.3 pc and obtained an $X_{\rm CO}$ factor about twice the Milky Way value using the virial theorem. Cloud velocity dispersions are a factor of $\sim3\times$ larger for their size than in the Milky Way in the starburst region, and surface densities are $\sim3\times$ higher as well, reflecting $\sim10\times$ higher pressures from cloud gravity.


\subsubsection{CO-Dark Gas} \label{sect-COdark}

An important uncertainty is the atomic or molecular state of the CO-dark gas that shows up in dust emission or other tracers in the envelopes of CO clouds. It takes about 1 magnitude of optical dust extinction to shield molecular gas from photodissociation (Lee et al. 2018, but see Hunt et al. 2023), and for Milky Way metallicity, this magnitude corresponds to a gas surface density of $\sim20\;M_\odot$ pc$^{-2}$. At lower metallicity, the dust column for one magnitude stays constant, but the gas surface density increases, making a relatively thick photon-dominated region (PDR) \citep{lee15,schruba18} with prominent \cii\ emission \citep{Cormier:19}. 

\begin{figure}[t!]
\includegraphics[width=3.75in]{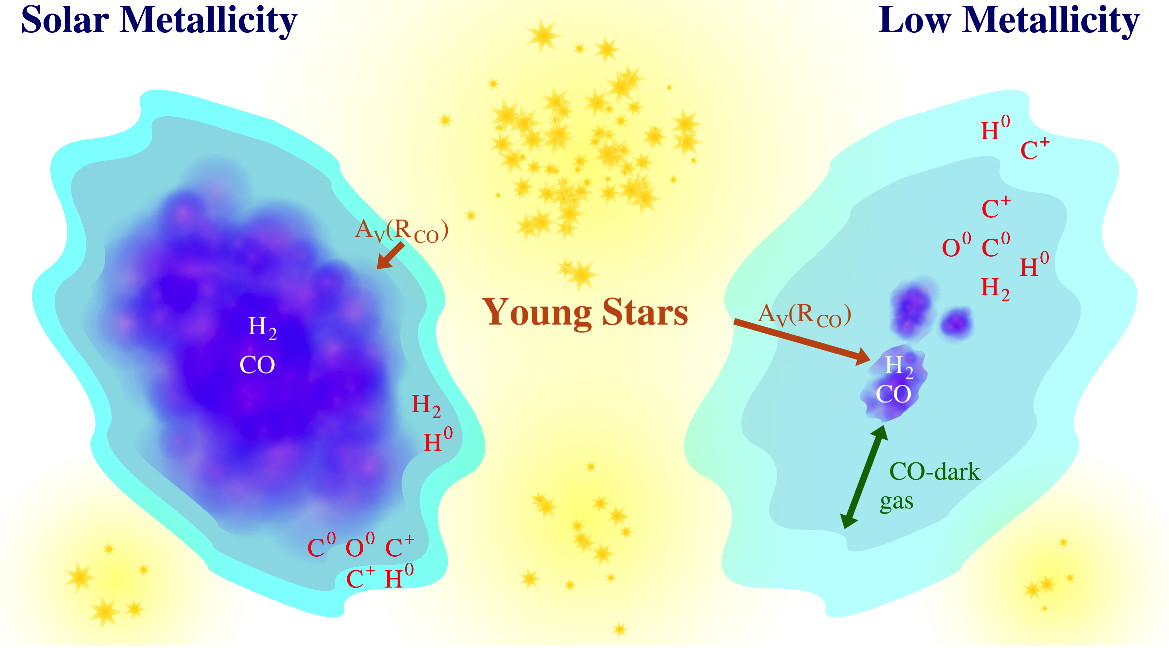}
\caption{Sketch of the differences between a molecular cloud
at solar metallicity (left) compared to one at low metallicity (right). The lower
dust and metal content of dIrrs results in a greater
dissociation of molecular gas and the formation of
a large envelope of CO-dark gas around small CO cores. The yellow region denotes ionized hydrogen. Figure adapted from \citet{madden20}. 
}
\label{fig-darkgas}
\end{figure}

For example, one can see the large PDR and tiny CO cores in WLM, a Local Group dIrr with an oxygen 
abundance of $0.13Z_\odot$ \citep[see \textbf{Figure \ref{fig-wlmse}},][see also Cormier et al.\ 2014]{Rubio:15}.
The expected reservoir of CO-dark gas at low metallicity compared
to a molecular cloud at solar metallicity is sketched in 
{\bf Figure \ref{fig-darkgas}}. 

Because the envelope is not easily observed in \HI, it is usually assumed to be H$_2$. However, if the mean density is low and feedback photo-dissociation is fast, as \citet{hu16,hu21} suggest in their models, or the dissociating radiation comes in through a few low-opacity sightlines \citep{seifried22}, then H$_2$ may not have time or opacity to form completely before it is destroyed by young stellar radiation. That would imply much of the peripheral gas is cold \HI\ and hard to detect in \HI\ emission line profiles. \cite{seifried22} suggest further that the CO-rich regions also have significant amounts of cold \HI, based on simulations of turbulent molecular clouds with time-dependent chemistry and radiative transfer.  

\HI\ line profiles in dwarf galaxies have been fit to cool and warm components without finding much cool \HI\ (see Section \ref{sect-coldhi}).
For example, \cite{koch21} suggested that optically thick (``top hat'') profiles in M31 and M33 result from multiple components and not cold \HI\ self-absorption. There is little systematic evidence for \HI\ self-absorption from local cold \HI\ either, i.e., surrounding molecular clouds in the solar neighborhood \citep{reach17,murray18}. An exception is in \cite{hayashi19}, who suggest on the basis of total gas observations through $\gamma$-ray emission that local CO-dark gas may in fact be optically-thick \HI. For the LMC, the ratio of cold \HI\ from self-absorption to total cloud mass, using $X_{\rm CO}=4\times10^{20}$ cm$^{-2}$ (K km s$^{-1}$)$^{-1}$ to get the cloud mass from CO, is only $\sim10^{-3}$ \citep{liu19}. 

Using a different method, \cite{togi16} found that the H$_2$ excitation temperature is reasonably approximated by a power law and then NIR observations of H$_2$ emission, which sample only the high-excitation lines, can be extrapolated to total H$_2$ content. With this method applied to 5 dIrrs, they found agreement between the extrapolated H$_2$ and the total gas masses inferred from dust emission and SFRs, which implies the CO-dark envelopes are H$_2$. 

Evidently, the molecular state of the CO-dark envelopes in dwarf galaxies is uncertain. Possibly, the H$_2$ fraction changes with spatial scale \citep{seifried22}, being lower for poorly resolved CO envelopes in dwarf galaxies than highly resolved CO envelopes in the local ISM. Alternatively, the H$_2$ formation rate could be higher than assumed in the simulations as a result of additional processes on grain surfaces \citep{wolfire22} or a larger dynamic range for density.


\subsection{Molecular Conversion Factor} 
\label{sect-conversion}

In the Milky Way CO is used as a tracer of the $H_2$ which makes up the bulk of molecular clouds. A conversion factor has been determined which, when applied to the CO measurement, gives the total molecular gas mass. Unfortunately, this conversion factor is expected to be a function of metallicity and possibly other environmental factors, and has been very hard to pin down for dwarf galaxies. Here we discuss the methods for determining the conversion factor at low metallicity and the problems with those values.

\subsubsection{Methods}\label{sect-conversion-methods}

As illustrated in Section \ref{sect-molecular-lowest}, various strategies have been developed to quantify the reservoir of CO-dark gas. One method assumes the observed CO represents the whole cloud and the mass is given by the virial theorem for the observed size and velocity dispersion. This virial mass is then divided by the luminosity to produce $\alpha_{\rm CO}$ \citep{wilson95}. This method cannot determine the envelope mass if it is not observed in CO and the envelope radius and velocity dispersion are not known. However, one might be able to assume that the CO cloud boundary pressure is determined by the weight of the overlying (invisible) envelope and derive the envelope column density from that. Contributions to the virial theorem from magnetic energy should be included too.

In another method, the dust mass in a cloud is derived from a fit of dust temperature, emissivity, and luminosity to multi-band FIR observations, and combined with a DGR previously calibrated for the appropriate metallicity from other observations.  The observed cloud is presumably defined by the region of high FIR emission, so the ratio of the dust mass in this region to the DGR is the total gas mass in the cloud. Then the \HI\ and ionized gas masses should be subtracted from the total cloud mass to determine the molecular part, which would include the CO part in the core. 

There are several uncertainties with the dust mass method. The dust emissivity depends on the grain composition, size, and temperature distributions, which may vary between dense and diffuse gas, and the fit to dust mass depends on the completeness of the SED and the applied SED model. The DGR depends linearly on metallicity for values near the Milky Way, but becomes non-linear at low metallicity \citep[see also Section \ref{sect-dusttogas}, and][for issues associated with dust-based estimates of total \hmol]{meixner13,jameson16, jameson18}. DGR measurements done elsewhere also have uncertainties about dark gas, and the DGR may vary from place to place in a galaxy \citep[e.g.,][]{hu23}. 

Alternatively, one can use \cii\ to infer $\alpha_{\rm CO}$. \cii\ comes from PDRs where far-ultraviolet (FUV) photons ionize weakly bound atoms like Carbon, which has an ionization potential of 11.3 eV \citep{nordon16}.  In a typical cloud, the CO core that traces the densest H$_2$ is surrounded by a PDR, which grows in size with decreasing metallicity \citep[e.g.][]{Bolatto:13}. Early KAO observations of the \ciiline\ line in IC 10 and the LMC found high \ciico\ ratios, suggesting substantial CO-dark gas \citep{poglitsch95, madden97, israel11}.  \hers\ and SOFIA extended these results, highlighting the diagnostic capabilities of \cii\ as a tracer of CO-dark gas in the LMC, SMC \citep[e.g.,][]{requena-torres16, jameson18, lebouteiller19, chevance20} and NGC 4214 \citep{fahrion17}. With \hers\ \cii\ and ALMA CO observations of the SMC, \citet{jameson18} found that CO emission with the Milky Way value of $\alpha_{\rm CO}$ traces only 5\% -- 60\% of the gas associated with CO; CO/\cii\ increases with \Av, confirming that \cii\ emission extends beyond the CO cores. 
 
The \hers\ Dwarf Galaxy Survey \citep[DGS;][]{madden13} observed a suite of FIR and MIR fine structure emission lines, including \cii, in 50 star-forming dwarf galaxies. These observations allowed detailed PDR and photoionization modeling \citep{cormier15, Cormier:19}, setting up a strategy to quantify the total dark gas using \cii. For example, 
\cite{Cormier:19} showed that the PDR covering factor increases with metallicity, and \cii\ is mainly in the PDRs. \cite{madden20} and \cite{ramambason23} used this strategy to conclude that CO traces a small fraction of the star-forming gas in dwarf galaxies, while \cii\ traces more than 70\% of the total molecular gas, most of it being the CO-dark gas. 

As for the dust method, there is an uncertainty with the \cii\ method in that the emission can come from diffuse \HI\ and ionized gas as well as the envelopes of CO cores \citep{ramambason23}.  If particular clouds are to be measured, such as the clouds associated with CO, then maps of the dust or \cii\ concentrations are needed to define the cloud boundaries. If the total dark gas mass is desired instead, i.e., regardless of association with CO or star formation, then the total mass from dust minus the mass of the CO cores derived from the Milky Way value of $\alpha_{\rm CO}$ should give about the same result as the total mass from \cii, which is presumably not present in CO cores because of their high opacity to FUV. The \cii\ method is currently used for high-$z$ galaxies to infer the total \hmol\ mass \citep[e.g.,][]{zanella18}.

A related method is to observe carbon in all of its important forms, i.e., \ci, \cii\ and CO, and to use the abundance of carbon relative to hydrogen to get the total hydrogen mass. Subtracting the \HI\ mass then gives H$_2$. \cite{pineda17} do this for the Magellanic Clouds and find substantial amounts of H$_2$ without CO, and that in H$_2$-dominant regions, most of the carbon is in the form of \cii. 

\subsubsection{Observations}\label{sect-conversion-observations}

CO-dark molecular envelopes have been observed in dust emission,
\cii, 
and \ci,
and they have been inferred from the virial theorem using the size and CO linewidth in the region. 
For a resolved CO cloud, the mass derived by these methods divided by the CO luminosity should be similar to the $\alpha_{\rm CO}$ of the Milky Way \citep{rubio93a}, 
i.e., independent of metallicity down to at least the SMC value \citep{bolatto08}. This similarity arises because the CO-dark envelope is mostly avoided when the CO cloud core is measured directly, and CO cores at low metallicity resemble CO clouds in the Milky Way (Section \ref{sect-molecular-lowest}).


Recent observations of the increase in $\alpha_{\rm CO}$ with decreasing metallicity for local dwarf galaxies are in \cite{leroy11}, \cite{schruba12}, \cite{hunt15}, \cite{amorin16}, \cite{shi16}, \cite{accurso17}, \cite{hunt20}, \cite{madden20}, \cite{hunt23} and \cite{ramambason23}.  Similar studies at high redshift show the same inverse relationship, although not generally to such low metallicities \citep{genzel12}. Of these studies, only \cite{hunt15,hunt20}, \cite{shi16} and \cite{ramambason23} apply their modeling to observations of galaxies with metallicities as low as $\sim$ 0.1 \zsol. Considering the small number of such galaxies with CO detections, the scaling of $\alpha_{\rm CO}(Z)$ is not well constrained \cite[see review in][]{Bolatto:13}.

Using the DGS observations and single {\it Cloudy} models, \cite{madden20} found an $\alpha_{\rm CO}$ well-correlated with metallicity and very steeply rising toward low $Z$. 
Also using the DGS data but with a statistical Bayesian framework to account for the structure of the unresolved gas with multiple components and a clumpy distribution of CO, \cite{ramambason23} found a wide scatter of $\alpha_{CO}$ that is correlated with CO clumpiness also. A steep relationship for $\alpha_{CO}(Z)$ corresponds to a more diffuse CO source while a flatter relationship found for some dwarf galaxies corresponds  to more clumpy CO structures. 

Submm transitions of \ci\ have also been used as tracers of CO-dark gas in observations as well as theory and simulations \citep[e.g.,][]{papadopoulos04, offner14, li18,jiao19}. \cite{glover16, heintz20} and \cite{ramambason23} show that \ci\ is an excellent tracer of total \mhmol\ in their models and that the conversion factor depends on metallicity approximately as $Z^{-1}$.

\subsubsection{Theory}\label{sect-conversion-theory}

Theoretical discussions and models of $\alpha_{\rm CO}(Z)$  in \cite{wolfire10} and \cite{krumholz11} 
consider uniform spherical clouds with dust extinction and grain-surface formation of \hmol\ 
to derive an envelope opacity and from this the CO-to-\hmol\ ratio. 
The resulting fractional abundance of CO depends on the fractional abundance of \hmol.
The strongest scaling with metallicity is through an 
$A_{\rm V}$ term, since $\alpha_{\rm CO}\propto\exp(2.12/A_{\rm V})$ in their model for $A_{\rm V}\propto Z$. 
A magneto-hydrodynamical simulation of H$_2$ and CO formation by Glover \& MacLow (2011) finds a comparably strong scaling of 
$\alpha_{\rm CO}\propto A_{\rm V}^{-3.5}$ down to $Z/Z_\odot=0.03$. 
These results agree with observations in \citet{hunt15} down to $Z/Z_\odot=0.1$ but appear too steep below that. 
Additional hydrodynamic simulations are in \citet{shetty11}; a recent detailed non-equilibrium chemical model of the ISM is in \cite{katz22}.

\cite{bialy15} determined the relative abundances of \hmol, CO and other important molecules from a detailed chemical network. 
\cite{bialy16} then examined dust opacity at the molecular transition, which is where the rate of \hmol\ formation, 
typically on dust, balances the \hmol\ photo-dissociation rate in the Lyman Werner bands. 
Because the radiation field is attenuated by both \hmol\ photo-dissociation and dust, metallicity plays a role in determining the fraction of the incident photons that dissociate \hmol. 
The mass surface density of the shielding layer has an additional dependence on the inverse of metallicity from the DGR, making the \HI\ envelopes of the \hmol\ regions even more massive at lower metallicity. \cite{schruba18} found good agreement between these models and observations of \HI\ in metal-poor galaxies.

\cite{hu16} modeled cold and warm \HI\ and cold H$_2$ in a whole dwarf galaxy with metallicity $Z/Z_\odot=0.1$. They found large H$_2$ fractions, 25\% to 70\%, in diffuse (non-star-forming) gas for the models with feedback, and small H$_2$ fractions for these models in the star-forming gas, defined to be where the density exceeds 100 cm$^{-3}$. 
The cold gas lies 3 to 10 times closer to the midplane than the warm gas because of its higher density and radiative shielding closer to the plane. Much of the gas with density $>100$ cm$^{-3}$ is cold \HI\ because feedback disrupts the H$_2$ in their models. 
In further studies, \cite{hu21} simulated the ISM with star formation in collapsing clouds for a galaxy patch 1 kpc square and 10 kpc high using metallicities ranging from 0.1 to 3 $Z$\solar. They found that the SFR is independent of metallicity, and, like their previous result, the H$_2$ is often out of equilibrium and at low relative abundance because of its long formation time. CO is less affected by the low H$_2$ abundance, however, making the H$_2$-to-CO ratio low in the cloud envelopes. 
Whereas the H$_2$ fraction was high in the CO regions, it decreased to low values, $<0.1$, in the CO-free molecular cloud periphery. 
At $Z/Z_\odot=0.1$, more than 40\% of the \hmol\ was in peripheral cloud gas with minimal CO and an \hmol\ fraction less than 10\%. \cite{hu22} discussed these results explicitly in terms of the $X_{\rm CO}$ factor: on large scales surrounding CO cores, molecular H$_2$ exists but with lower $X_{\rm CO}$ than calculated from equilibrium models, and on the small scale of the CO region, dust opacity is high enough to make the \hmol\ fraction high regardless of metallicity, and then $X_{\rm CO}$ approaches the Milky Way value. 

\cite{hirashita17} and \cite{hu23} calculated $X_{\rm CO}(Z)$ for various metallicities, including dust evolution in the DGR. While Hu et al.\ continue to model non-equilibrium chemistry, \cite{hirashita17} note that the H$_2$ formation time can be less than the free fall time for densities exceeding $\sim10^3$ cm$^{-3}$, and thus H$_2$ is assumed to reach equilibrium in their models. This difference arises because the formation time scales inversely with density for a given DGR, being a collisional process, while the free fall time scales inversely with the square root of density, driven by self-gravity.   Both studies conclude that molecule formation can be enhanced in local regions of high DGR, which may arise through dust growth or dust formation in supernovae and evolved stars. \cite{hirashita17} use the chemistry in \cite{glover11} modified for a different dust abundance and grain size distribution that was calculated explicitly and found good agreement with the observed $\alpha_{\rm CO}(Z)$ down to 20\% solar abundance. \cite{hu23} also calculated dust evolution but with a time-dependent chemistry in a hydrodynamic model of a WLM-like dwarf galaxy with 10\% solar abundance. Hu et al.\ found that dust growth in dense gas is required to reproduce the observed CO luminosity at the observed SFR; otherwise UV radiation from star formation will destroy the CO clouds. Thus the DGR is higher in CO clouds than in proportion to the overall metallicity. Hu et al.\ continued to get a low H$_2$ fraction because of its slow formation, and so essentially found no CO-dark molecular gas, which made $\alpha_{\rm CO}$ comparable to that in the Milky Way. 
    
\cite{accurso17} and \cite{hunt20} make the point that $\alpha_{\rm CO}$ depends on both the sSFR through variations in the molecular depletion time, $\tau_{\rm d}$, and the metallicity, which is partly determined by molecular photo-dissociation in interclump gas shielded to various degrees by dust (Israel 1997). 
Higher sSFR leads to lower $\tau_{\rm d}$ 
\citep{huang14}
although there is no strong correlation for local galaxies in \cite{jameson16}, and lower $L^\prime_{\rm CO}/SFR$, 
at solar and higher metallicity, when molecular clouds are presumably saturated with CO. Here $L^\prime_{\rm CO}$ refers to a measurement in observational units, K km s$^{-1}$ pc$^2$ rather than $L_\odot$.

\subsection{Molecular Gas in Tidal Dwarfs and Stripped Tails} \label{sect-molecular-tidal}

Tidal dwarf galaxies (TDGs) form as self-gravitating debris of gas and stars in tidal tails \citep[see review in][]{duc13}. Because the gas comes from a larger galaxy, it has the higher metallicity of the larger galaxy and therefore a high metallicity for its mass. As a result, CO is easier to detect in TDGs than in dIrrs of similar mass.

The CO luminous and virial masses in a TDG that appears to have been stripped off of NGC 3077 from an interaction with M81 exceed $10^7\;M_\odot$ in each of three complexes; the CO exists beyond where the \hi\ column density peaks at $1.5\times10^{21}$ cm$^{-2}$ \citep{heithausen00}. This peak exceeds the \hi\ to H$_2$ threshold in the solar neighborhood, where it is $\sim5\times10^{20}$ cm$^{-2}$ (Savage et al. 1977; however see Seifried et al.\ 2022). 

\cite{braine01} discovered CO emission centered on the \HI\ peaks in 8 TDGs
and noted that the CO luminosities are $\sim100\times$ higher than in dIrr galaxies at the same stellar luminosities because of the 
$\sim1/3Z$\solar metallicities in the TDGs. In the TDG near NGC 3077, the SFR is low for the molecular mass, but in the TDGs studied 
by \cite{braine01}, it is about the same as in the Milky Way. \cite{brinks04} observed \HI\ and CO at 750 pc resolution in the TDG near Arp 245 and concluded it is gravitationally bound.

CO has been observed in many other TDGs as well: in the intergalactic region of Stephan's Quintet
and VCC 2062 \citep[][and references therein]{lisenfeld16},
near VV 114 \citep{saito15}, near M82 \citep{pasha21}, and in tidal clumps in the Leo ring \citep{corbelli23}. 
ALMA observations of Arp 94 resolved 111 CO(2-1) clouds in the TDG down to 45 pc resolution with a mass spectrum similar to that in the Milky Way \citep[][and references therein]{querejeta21}; the molecular fraction is high ($\sim50$\%) and the cloud velocity dispersion is higher for a given cloud mass than in the Milky Way, although the CO clouds look fairly normal. 

Ram-pressure swept tails of galaxies may contain CO also, as in ESO137-001 where ``fireballs'' aligned with the tail could have formed molecules {\it in situ} \citep{jachym19}. The isolated intra-cluster cloud in Abel 1367, which may be the remnant of a tidal tail, contains over $10^8\;M_\odot$ of molecules, mostly not star-forming \citep{jachym22}.

\section{DUST AND METALLICITY} \label{sect-dustandmetals}

One of the most prominent characteristics of dwarf galaxies is their low abundance of metals and dust. 
Metals originate in stars, with dust forming in novae and supernovae \citep{haenecour19,hoppe22}, in the expanding atmospheres of evolved stars \citep[see review in][]{dwek05}, and more generally in the denser regions of the ISM. 
Dust is important for interstellar molecules, as it is the primary site for \hmol\ formation 
while shielding molecular clouds from photodissociative radiation \citep{bialy15}.

There is spectroscopic evidence that the composition of dust in dwarf galaxies is made up of the main Galactic dust components, silicates \citep[e.g.]{thuan99, jones23}, carbonaceous dust including polycyclic aromatic hydrocarbons (PAHs) (Sect. \ref{sect-dustprops}) \citep[e.g.]{madden06}. Although the proportion and size distribution of these components differ \citep[e.g.,][]{zubko04, jones17, hensley23}, we can reasonably use Milky Way dust models to study low metallicity galaxies. Dust properties depend on gas density and proximity to radiation, which determine the rate of condensation and sublimation at the surface, and on dynamical processes such as shocks, which can sputter or shatter the grains in collisions \citep[see review in][]{galliano21}.

The chemical elements we measure in a galaxy today serve as a history of the evolution of its stellar populations, which, through the process of nucleosynthesis, are the sources of the elements that are injected into the ISM and cycled through successive generations. 
The chemical history of dwarf galaxies is also linked 
to the history of accretion (Section \ref{sect-accretion}) and outflow \citep[Section \ref{sect-atomic-holes}, e.g.,][]{mccormick18}. Metallicity increases with stellar mass approximately as $M_{\rm star}^{0.3}$ near $M_{\rm star}\sim10^{10}\;M_\odot$ with a shallower slope for more massive galaxies \citep{kewley08,kirby13}. The reasons for this increase include a greater resistance to gas escape \citep{tremonti04} and a higher efficiency of star formation \citep{calura09} at higher galaxy mass.  As stars produce metals, the relationship evolves over time \citep{torrey19}; for dwarfs, metallicity increases by up to 0.1 dex at constant mass from $z=3$ to 2 \citep{li22}.

\subsection{Heavy elements in the gas and dusty phases} \label{sect-heavyelements}

While many techniques exist to estimate the metallicity of galaxies \citep[see review by][and references within]{maiolino19}, most of the metallicities of gas-rich dwarf galaxies in the local universe and out to moderate redshifts have been measured in the ionized gas phase, frequently via the oxygen nebular lines (expressed as 12 + log(O/H)) \citep[e.g.,][]{skillman89, izotov00, kunth00}.  Oxygen is one of the most abundant heavy elements.
The ``direct" metallicity method measures the strong optical [OIII] lines ($\lambda$4363/$\lambda$5007) and compares these to calibrations based on lines that are sensitive to electron-temperature \citep[e.g.,][]{pettini04}. Oxygen is the standard against which the abundances of other elements (Fe, C and N in \HII\ regions)  are compared \citep{pagel03} which are all mainly produced in massive stars. The H abundances in the O/H ratio are determined often by optical hydrogen recombination lines. Studying the elemental abundance ratios and how the patterns vary as a function of the metallicity in dwarf galaxies can provide a window into the chemical evolution of early galaxy environments. 

\begin{marginnote}[]
\entry{Metallicity}{Metallicity, proportion of atoms heavier than H and He, is often measured in the gas phase using optical oxygen nebular lines and calculated as 12+log(O/H).}
\end{marginnote}

Understanding the origin of N and the N/O ratio in dwarf galaxies has been a persistent problem.  At low metallicities, N/O is observed to be relatively constant with $Z$  but often with a wide scatter \citep[e.g.,][]{kobulnicky96, izotov99, pilyugin03, spite05, lopezsanchez10, roy21}, and with some evidence for an increase of N/O in  extremely low-$Z$ galaxies \citep[e.g.,][]{guseva11}. While O is a primary element (synthesized from the initial stellar hydrogen and helium), the chemical origin of N, and whether it is a primary or secondary element (production in previous stellar generations) at low $Z$, has been challenging to model. 
Preference exists for primary N production in metal-poor stars  \cite[e.g.,][]{spite05, roy21}.  High mass stars are the dominant producers of oxygen. Nitrogen in dwarf galaxies has been attributed to production in massive stars \citep[e.g.,][]{izotov99} while other studies suggest low to intermediate mass stars \citep[e.g.,][]{skillman98, henry00, spite05}. Recent models by \citet{roy21} suggest  that dwarf galaxies retain their massive stellar winds, thus producing the observed N/O vs O/H relation, and that retention of supernovae ejecta is not likely to account for the observed N/O until higher metallicities are reached and AGB stars contribute secondary N.

The EMPRESS survey of extremely metal-poor galaxies (about 2\% \zsol) measured abundances of Ne/O and Ar/O to be  comparable to those of other local dwarf galaxies with lower than expected N/O \citep{kojima21}, consistent with findings by \citet{guseva11}. However, in contrast, the Fe/O is excessively high: 90\% \zsol\ to 140\% \zsol\ proposed by \citet{kojima21} to originate from supermassive stars,
while \citet{guseva11} show a prominent decrease in Fe/O as metallicity increases, likely due to the depletion of Fe onto dust grains.

Metallicity variations measured in the gas phase across dwarf galaxies are, for the most part, negligible \citep[e.g.,][]{kobulnicky96, croxall09} unlike the metallicity gradients across large star-forming disk galaxies 
\citep[e.g.,][]{sanders12}.
Some elemental abundances in the neutral \HI\ gas of I Zw 18,  
one of the most metal-poor galaxies in the local universe, were observed with UV absorption lines to be lower ($\sim$1/46\zsol) than those in \HII\ regions ($\sim$1/31\zsol), measured with the commonly-used optical lines \citep{aloisi03, lebouteiller13}. One explanation surmises the infall of more metal-poor gas mixing with the outer \HI.  Cosmological simulations of dwarf galaxies have explained the uniformity of metallicity as the result of turbulent diffusion \citep[e.g.,][]{williamson16, escala18}. However simulations following the chemical evolution of each enrichment event  for individual elements indicate that metal mixing is not uniform across a species \citep{emerick18}.
 
The depletion of elements from the gas phase to dust modifies the relative  gas phase elemental abundances and makes it challenging to get an accurate assessment  of them \citep[e.g.,][and references therein]{jenkins09, tchernyshyov15, decia16, roman-duval21}. Depletion  tells us  more about the dust properties.  Elemental depletion patterns of the LMC ($Z=0.5$\zsol) and SMC ($Z=0.2$\zsol) have been determined from UV absorption measurements of metal ions and analysed along with stellar abundances in studies by 
\citet{tchernyshyov15}, \cite{jenkins17}, \cite{roman-duval22}, and \cite{ konstantopoulou22}. While gas-phase abundances vary throughout the galaxies, depletion patterns are not very different from those of the Milky Way.   
As a result, dust properties in the LMC  are similar  to abundance-scaled dust  properties in the Milky Way, unlike those in the SMC, which does not scale similarly to the Milky Way but has wider variations \cite[see comparison of abundances and depletions in the LMC, SMC and Milky Way in][]{galliano18}. 
Depletion studies show that while star formation histories can differ widely in galaxies, the origin of dust in low metallicity galaxies and more metal-rich galaxies is similar \cite [e.g.][]{konstantopoulou22}.

\subsection{Dust Properties of Dwarf Galaxies} \label{sect-dustprops}
The dust properties of galaxies, such as the dust mass relative to gas and the distributions of temperature, composition, and size, can be extracted from models of the dust spectral energy distributions (SEDs).  The specific dust characteristics depend on the approach adopted for the macrophysics of the SED modeling (e.g. the radiative transfer), and the microphysics in the dust models (composition). Modified black body (MBB) models, assuming a single dust temperature, T$_{\rm dust}$, are often used to interpret the observed SEDs of the dust emission. These models relate T$_{\rm dust}$ to the dust mass, 
and the dust emissivity index, $\beta$, which is dependent on the dust composition.
The MBB models, while conveniently simple, are found to produce lower dust masses \citep[e.g.][]{dale12, remy13} than more physically realistic models \citep[e.g.][]{dale12, remy15, galliano18b, hunt19, aniano20}, which assume a starlight intensity distribution. With a distribution of radiation fields as is the case in galaxies (hence of equilibrium grain temperatures), application of a MBB results in a $\beta$ that is artificially decreased to compensate for the broadening of the SED, which leads to a higher fitted temperature, and thus to a lower dust mass
\citep[see][and references within for a detailed discussion of the phenomenology of SED modelling and dust models]{galliano18}.

\iras\ brought us the first 12 to 100 \mic\ view of the global dust properties of dwarf galaxies, highlighting the trend of warmer overall temperatures compared to massive spirals, as determined by the 60 \mic/100 \mic\ ratio. These higher temperatures resulted in lower inferred dust-to-gas ratios (DGRs) 
\citep[e.g.][and references therein]{melisse94,hunter01}.
Early spectroscopic observations of NGC 5253 and II Zw 40 also revealed an absence of PAHs \citep{roche91}, which are carbonaceous dust grains with prominent MIR emission features. In contrast, PAHs are pervasive in star-forming disk galaxies containing PDRs \citep[][and references within]{tielens08}. With greater sensitivity, higher spatial resolution and more MIR to FIR wavelength coverage in larger samples of dwarf galaxies, \iso\ and then \spitz\ confirmed these characteristics of low metallicity galaxies, and established more clearly the paucity of PAHs in dwarf galaxies \citep{reach00, houck04, galliano05, engelbracht05, madden06, draine07, hunter07, galliano08, rosenberg08, hunt10, sandstrom10, chastenet19, aniano20}. 
As PAHs are suggested to be the main carriers of the 2175 \AA\ UV bump in the dust extinction curve \citep{fitzpatrick86, mathis94, misselt99, li01}, 
the Large Magellanic Cloud, with its relatively low PAH abundance, shows a lower  2175 \AA\ bump than the Milky Way, and the SMC, with fewer PAHs than the LMC, has an even lower 2175 \AA\ feature \citep[e.g.][]{prevot84, fitzpatrick86, gordon98, weingartner01, li02, draine03c, gordon03, fitzpatrick07, hirashita20b}.


High spatial-resolution FIR-to-submm observations arrived with the \hers\ Space Observatory, thus going beyond the FIR peak into submm wavelengths where the total dust mass can be more accurately quantified.  \hers\ carried out the Dwarf Galaxy Survey \citep[DGS;][]{madden13}, observing 50 dwarf galaxies with a wide range of metallicities, including 2 galaxies that are among the most metal-poor in the local universe, I Zw 18 (0.05\zsol) and SBS0335-052 (0.03\zsol). 
Comparison of these dwarf galaxies with the more metal-rich sample from the \hers\ KINGFISH survey  \citep{kennicutt11, dale12, dale17}, and subsequently the DustPedia survey \citep[e.g.][]{galliano21}, provided a broad range of environments and star-formation activities to help us understand how dust properties evolve as galaxies become enriched over almost 2 dex in metallicity \citep{remy13, remy15, galliano21}.

Compared to their gas and stellar masses, the dust masses in dwarf galaxies are lower than those of more metal-rich galaxies. Stellar feedback in dwarf galaxies carves out ISM channels \citep[e.g.][]{emerick19}, making more porous ISM structures (Section \ref{sect-atomic-holes}), as also evidenced by high galaxy-wide \lfir\ and harder radiation field revealed by the presence of high-energy fine structure lines emitting in the MIR and FIR \citep[e.g.][]{cormier15}.   
Consequently, combined with the smaller grain size distribution, the overall dust temperatures in dwarf galaxies are warmer, as evidenced by their dust SEDs peaking at shorter wavelengths, often between $\sim$ 40 to 60 \mic, compared to 100 \mic\ and beyond for dustier galaxies \citep[e.g.][]{hirashita08, remy13, galliano21}. Overall, dwarf galaxies show broader profiles around the peaks of their SEDs \citep{remy15} reflecting a wider temperature distribution for the grains.

\subsection{Low metallicity and dust size effects} \label{sect-dustsize}

SED modeling suggests that the dust size distributions in dwarf galaxies can shift to smaller sizes compared to more metal-rich galaxies \citep[e.g.,][]{galliano05}, although degeneracies exist in the SED approach. The SMC extinction curve also suggests a shift to smaller sizes \citep[e.g.][]{cartledge05}.  
 At extremely low metallicities (e.g.\ I Zw 18), theoretical models by \citet{hirashita20} predict large grains ($\sim$ 1 \mic) as the mass is dominated by grains formed in SN ejecta. At more moderate metallicity (e.g. SMC), grains are shattered via shocks and become smaller, down to a few nm \citep[e.g.][]{jones04, slavin20, priestley22b}, while at higher metallicity they can become larger as they accrete gas-phase species in the ISM.  The dust size distribution, biased toward smaller grains, tends to increase the dust temperature distribution. The predicted increase of the silicate-to-carbon dust ratio via shock processing and differing production time scales contribute to the characteristically different extinction properties and dust SEDs observed in dwarf galaxies. The evolution of the dust size distribution can also affect the metallicity dependence of the CO-to-\hmol\ conversion factor   \citep[Section \ref{sect-conversion};][]{hirashita23}.

PAHs are often associated with UV-illuminated edges of PDRs around massive stars.
The lower abundance of PAHs in dwarf galaxies can therefore be a consequence of low filling factors for PDRs (see Section \ref{sect-COdark}). On the other hand, the low abundance of PAHs can be attributed to lower dust shielding, encouraging PAH destruction from hard radiation environments \citep[e.g.][]{madden06, hunt10} and supernova shocks \citep[e.g.][]{ohalloran06, micelotta10}. An example of these combined effects observed in the SEDs of star-forming dwarf galaxies is SBS0335-052, which is one of the lowest metallicity galaxies in the local universe ($\sim$1/40 \zsol). SBS0335-052 has MIR silicate absorption features on a continuum without PAH features, and an SED peaking at $\sim$ 30 \mic, an unusually short wavelength for the FIR peak, indicating very warm dust \citep{thuan99,houck04}.

The fraction of total dust mass in the form of PAHs correlates with metallicity at 12 + log O/H $<$ 8.0 \citep[e.g.][]{draine07b, remy15, aniano20}.  With a sample of $\sim800$ galaxies, \cite{galliano21} found this correlation to be striking over 2 dex in metallicities and also found a weaker anti-correlation with the radiation field. The anti-correlation may be related to the destruction of PAHs via photoprocessing in the hard radiation field of the diffuse ISM, which also goes hand-in-hand with the overall warmer dust and the presence of very small grains (few nm) emitting in the MIR continuum \citep[e.g.][]{madden06, rosenberg08, hunt10, sandstrom10, chastenet19, aniano20, hirashita20, galliano21}.  For example, \cite{sandstrom10} and \cite{chastenet19} proposed that in the Magellanic Clouds, molecular clouds are the sites where PAHs form before photoprocessing in the diffuse ISM. Another PAH evolution scenario has them being produced by the shattering of large carbon grains \citep{seok14}. Moreover, PAHs may not yet have been produced in the lowest metallicity galaxies due to a delayed production of carbon stardust \citep{dwek05, galliano08}.

\subsection{Dust-to-gas mass ratios} \label{sect-dusttogas}

Measurements of the DGR in galaxies over a wide range of metallicities give us some insight into how dust may have evolved from early, metal-poor galaxies into metal-rich galaxies, such as our Milky Way. Interpretation requires a dust evolution model, dust and gas mass measurements and metallicity measurements.  While the number of very low metallicity dwarf galaxies with measured dust masses is rather limited due to telescope sensitivity issues \citep[e.g. SBS0335-052,][]{hunt14, remy15, cormier17}, the DGR has been measured in galaxies with metallicities covering almost 2 orders of magnitude
\citep[e.g.][]{remy15, devis19, galliano21, cigan21}.

\begin{figure}[t!]
\includegraphics[width=3.25in]{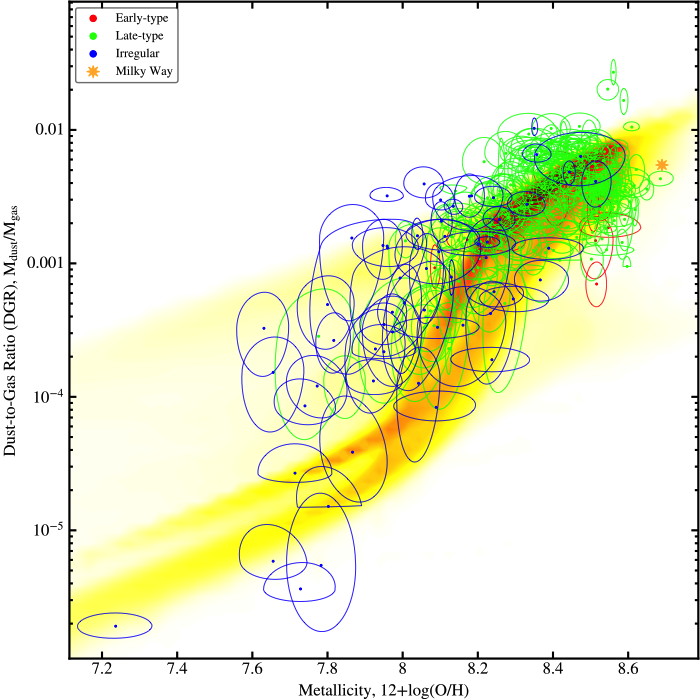}
\caption{Dust-to-gas mass ratio (DGR) as a function of
galaxy oxygen abundance, fitted with the dust evolution model of \citet{galliano21}. The skewed uncertainty ellipses are from the SED fitting of the DustPedia plus DGS galaxy samples. The data are color-coded in the legend according to galaxy type. The Milky Way is represented by an orange star. The posterior probability distribution of the dust evolution model is shown as a spread in yellow-orange density. Figure is adapted from \citet{galliano21}.
For reference, solar oxygen abundance is 12 + log(O/H)=8.69 \citep{solaroxygen}.
}
\label{fig-dgr}
\end{figure}

{\bf Figure \ref{fig-dgr}} demonstrates the behavior of the DGR as a function of
metallicity. At very low metallicities, 12 + log(O/H) less than 7.9 (Z less than
0.2 \zsol), condensation dust in supernova ejecta appears to be the dominant
production process, but it has a low efficiency \citep[e.g.,][]{asano13, devis17, galliano21}. Dust is a very meager
component of the ISM at the lowest metallicities - low compared to the already low metal abundances in these galaxies, and lower still compared to the gas in
star-forming metal-rich galaxies. As the build-up of dust progresses, the growth of grains via accretion of metals from the gas phase takes place in the ISM \citep[e.g.,][]{asano13, zhukovska14,zhukovska16} and
the DGR begins an abrupt rise by about 2 orders of magnitude between 12 + log(O/H) $\sim$ 7.9 and 8.2. Above this ``critical metallicity", non-linear regime \citep{asano13}, an approximate proportionality of DGR with metallicity is observed, albeit with scatter attributed to different
star formation histories and different grain formation and destruction processes \cite[e.g.][]{galliano08, remy14, devis19, cigan21, galliano21}.

\subsection{Submillimeter excess} \label{sect-submm}
Beyond the FIR peak of galaxy SEDs, the observed submm emission is sometimes seen as an excess that is not explained by a combination of state-of-the-art SED models, synchrotron and free-free continua, molecular lines and CMB fluctuations. 
This excess was first reported in the Milky Way  by \cite{reach95} and shown to be present in some dwarf galaxies \citep[e.g.][]{lisenfeld02b, galliano05, galametz09, bot10, galametz11, galliano11, dale12, chang21}
as well as in solar-metallicity galaxies \citep[e.g.][]{kirkpatrick13}. 
\hers\ observations brought more attention to this excess, which seems to begin around 500 \mic. The presence of the  excess is model-dependent, showing up more often with MBB SED models \citep[e.g.][]{remy13, hermelo16, relano18, chang21}, but as well in full state-of-the-art SED models \citep[e.g.][and references within]{galliano18}.


The origin of this excess remains an open issue. One suggestion has been a very cold dust component (T $\sim$ 10 K), which implies a large dust mass and significant dust shielding to maintain the low temperature, unrealistic unless the cold dust existed in dense clumps \citep{lisenfeld02b,galliano05, cormier17}.   
For some cases, grain emissivity variations with temperature can explain the submm excess \citep{meny07, demyk22} or magnetic grains, which seem to explain the excess in the SMC \citep{draine12, bot10}. Spinning grains have also been proposed \citep{draine98b, bot10, draine12} as a contribution to the mm and submm emission \citep[see review by][]{dickinson18}, but cannot alone explain the submm excess. \citet{bot10} suggested that spinning grains should be combined with a temperature-dependent emissivity to explain the excess in the SMC. In the  moderately-low metallicity galaxies LMC and M33, the submm excess is shown to prefer the more diffuse ISM, rather than the molecular component \citep{galliano11, hermelo16, relano18}. Until more sensitive telescopes can survey a statistically significant number of metal-poor galaxies in  well-resolved FIR-mm continuum bands, the submm excess remains an outstanding question.

\subsection{Dust evolution: the story told by the dwarf galaxies} \label{sect-dustevol}
To understand the evolution of dust in galaxies throughout cosmic times, it is necessary to understand the metal enrichment processes from the lowest metallicity primeval galaxies to more chemically evolved galaxies, already rich in metals. Nailing down the dust properties in dwarf galaxies is essential to form a complete picture of how the buildup of dust transpires in galaxies. The diversity of local Universe dwarf galaxies with varying metal-abundances can be thought of as laboratories providing individual snapshots of different metal-enrichment stages which, taken together as an ensemble, can help to construct comprehensive self-consistent dust evolution models. Most dust evolution models since \citet{dwek80} are usually one-zone models following the global content of dust, metals, gas and stars as a function of time \citep[e.g.][]{dwek98, asano13, zhukovska16, devis17, delooze20, hirashita20, nanni20, triani20, galliano21}, but not all models are constrained by observations of the very low metallicity galaxies.  Dust formation and destruction models attempting to reproduce the observed dust mass and predict the extinction curve  should assume or model a dust size distribution as a function of time \citep[e.g.][]{asano13b, asano14, mckinnon18, hirashita20b, makiya22}.

\citet{galliano21} perform a hierarchical Bayesian fit of a one-zone dust evolution model to the observed stellar mass, gas mass, dust mass, metallicities and star formation rates of 556 galaxies of a wide range of galaxy types ({\bf Figure \ref{fig-dgr}}). Their model reproduces the ``critical metallicity" threshold above which grains begin to accrete metals and dust grows efficiently, thereby enriching the ISM. The very low metallicity galaxies of the sample (lower than Z=0.2 \zsol) are critical to constrain the full dust evolution model especially in the regime where dust production is dominated by SN II condensation. While most studies agree on the process of dust evolution at the lowest metallicities, some scenarios for the enrichment of the higher metallicity galaxies have supernovae as the net dust producers \citep[e..g.,][]{delooze20, nanni20}. Spatially resolved studies would help to improve on the existing degeneracies in dust evolution models, but will be challenging for the lowest metallicity galaxies, especially at FIR wavelengths.

\section{ENVIRONMENTAL EFFECTS}

Mergers, interactions, and gas accretion can change dIrr galaxies from quiescent, with generally low surface brightness and small star-forming regions that produce low-mass clusters, into highly active, with dense molecular clouds, high specific star formation rates and massive young star clusters that produce giant shells and possibly blow-out into the halo. The fraction of dIrrs that are recent mergers or current accretion sites is uncertain, but there are clear examples of each as determined using several types of evidence. A few examples are reviewed here.

\subsection{Mergers} \label{sect-mergers}
Mergers and accretion are important for galaxy growth at high redshift, and these processes are still present in dIrrs today, although with reduced frequency. \cite{zhang20} present observations of giant stellar shells surrounding the BCD galaxy VCC 848. Computer simulations show how shells like this are evidence for a merger of two smaller galaxies, which in the case of VC 848 started around 1 Gyr ago. Starbursts accompanied this merger, as seen in the age distribution of massive star clusters. \cite{paudel17} previously found old stellar shells indicative of mergers in three Virgo cluster early-type dwarfs, and \cite{chhatkuli23} studied 6 BCDs with giant stellar shells. Other types of evidence for mergers are halo stellar streams, as seen in And II by \cite{amorisco14}, extended stellar halos, as seen in Tucana II by \cite{chiti21}, steep or unusual metallicity gradients \citep{vanzee98,grossi20,taibi22}, non-planar inner and outer disks, as in Ark 18 \citep{egorova21}, off-center black holes \citep{reines20,bellovary21}, and double AGN nuclei \citep{micic23}. The dwarf merger Mrk 709 has an interconnecting bridge of massive star clusters and an AGN in one component \citep{kimbro21}. The dwarf galaxy NGC 4449 apparently has an even smaller galaxy falling in \citep{martinez12}. 


\cite{stierwalt15} compiled a catalog of 104 dwarf galaxy pairs and showed that the specific SFR increases by a factor of about 2 when the galaxies get closer than 50 kpc. \cite{pearson16} studied the 10 closest of these pairs and suggested that the interaction disperses the disk gas, which then reaccretes later. Mergers are not obvious for most dIrrs: \cite{higgs21} found no evidence for merger structures in isolated local dwarfs considering galaxies closer to the Milky Way than 3 Mpc and more distant than 300 kpc from either the Milky Way or M31. \cite{martin21} suggested, based on a cosmological simulation, that less than 20\% of dwarfs have morphological evidence for a merger at a redshift of $z=3$, and less than 5\% have merger evidence at $z=1$; overall, mergers account for only 10\% of the galaxy mass. Dwarf galaxy interactions are more common than mergers. \cite{paudel20} studied an interacting pair of dwarfs similar to the LMC/SMC pair near the spiral galaxy NGC 2998, which is like the Milky Way; they concluded that NGC 2998 is in a denser environment than the Milky Way and that the dwarfs near it are interacting primarily by tidal forces, whereas the LMC and SMC also show the effects of ram pressure stripping.

\subsection{Accretion} \label{sect-accretion}

\begin{figure}[t!]
\includegraphics[width=4.75in]{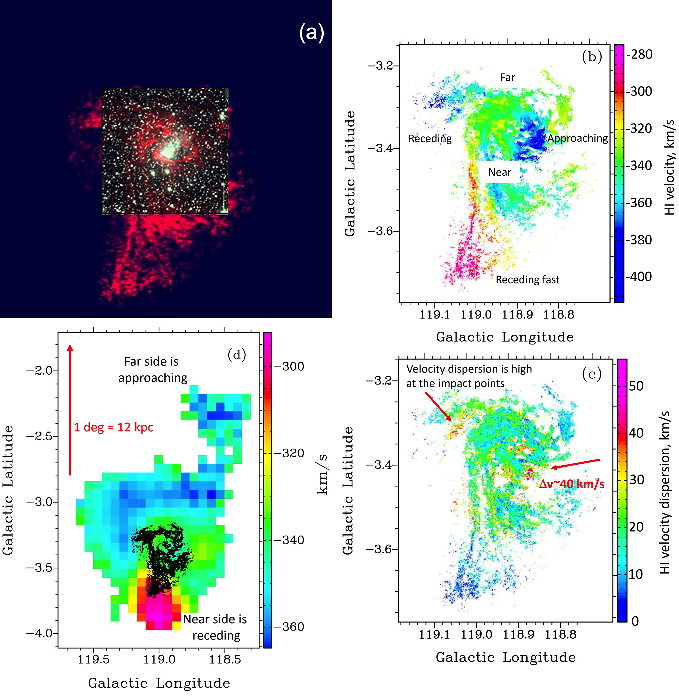}
\caption{Local Group dIrr IC 10 with accretion.
(a) Optical + JVLA-\HI\ image (in red) with low \HI\ contrast in the center of the panel, and the same \HI\ image with higher contrast in the outer parts; (b) \HI\ velocities with blue-shifted gas on the
far side and red-shifted gas on the near side, indicating accretion (the near and far sides are determined from the winding of the spiral arms); (c) a larger view 
of \HI\ from GBT with the same sense of accretion and the JVLA image in black; (d) \HI\ velocity dispersion suggesting
an increase in turbulent speed near the impact points.
From \citet{ashley14} with permission, reproduced by permission of the AAS.
}
\label{fig-IC10}
\end{figure}

Accretion onto dwarfs can be inferred in several ways, with direct evidence for gas flows being the most convincing. An example of this is in
IC 10, which is a dwarf galaxy at a distance of 0.7 Mpc  \citep{ashley14,namumba19}. 
{\bf Figure \ref{fig-IC10}a} from \citet[][]{ashley14} shows an optical image with \HI\ from the Karl G.\ Jansky Very Large Array superimposed and {\bf Figure \ref{fig-IC10}b} shows the \HI\ velocities with the near and far sides of the galaxy indicated along with the approaching and receding components of \HI. The nearside has gas receding from us, which means it is accreting onto the galaxy. A larger scale view from the Green Bank Telescope is in {\bf Figure \ref{fig-IC10}c} from \citet[][]{ashley14}, which shows \HI\ to much larger distances following the same sense of motion, i.e., approaching the galaxy on both the near (red color) and far (blue color) sides. These accretions apparently agitate the \HI\ in the disk also, as shown in {\bf Figure \ref{fig-IC10}d} which plots the \HI\ velocity dispersion. 
The dispersion is low in the southern accretion stream but high in two places that could be stream impact sites. The accretion rate can be determined from the \HI\ mass and infall speed. For the northern filament, $M(HI)=6\times10^5\;M_\odot$, the relative speed is $\Delta v=15$ km s$^{-1}$, and the stream length is $L\sim7$ kpc, which combine to give an accretion rate of $M\Delta v/L=0.001\;M_\odot$ yr$^{-1}$ and a time to finish of $\sim0.6$ Gyr. For the southern stream, $M(HI)=10^7\;M_\odot$, $\Delta v \sim 30$ km s$^{-1}$, and $L\sim7$ kpc, giving an accretion rate of $\sim0.04\;M_\odot$ yr$^{-1}$ and 0.2 Gyr to finish. The SFR of $0.04\;M_\odot$ yr$^{-1}$ \citep{ashley14} is comparable to the accretion rate in the southern stream.

Analogous cases might be DDO 53 \citep{egorov21} and Pisces A \citep{beale20}. The BCD galaxy NGC 2915 
has high-speed clouds with different metallicities and a counter-rotating gas disk with respect to the stars, which is also evidence for accretion \citep{tang22}. 
In NGC 1569 (see outflow discussion in  Sect. \ref{sect-atomic-holes}),
\cite{muhle05} observed an \HI\ hole in the center of the wind and extended \HI\ halo structure suggesting accretion, as found by \cite{stil02} and with deeper observations by \cite{johnson12}. 
Not all dwarfs are accreting though: NGC 6822 seems to have a net outflow \citep{fukagawa20}. The MHONGOOSE program at the MeerKAT radio telescope is designed to detect gas accretion onto nearby galaxies using faint \HI\ emission 
\citep{Sardone:21}.

Accretion to dwarfs may also be observed as off-center star formation hotspots, resembling ``tadpoles’’ \citep{vandenberg96}, especially if the star formation region has lower metallicity than the rest of the galaxy \citep{sanchez18,ju22,valle23}.  Three related studies of extremely metal-poor galaxies show metallicity drops at bright star-forming regions: \cite{sanchez13} observed 7 tadpoles and found 6 with metallicity drops, \cite{sanchez14} observed 7 and found 2 with metallicity drops, and \cite{sanchez15} observed 10 and found 9 with metallicity drops. CO observations of one of these, Kiso 5639 \citep{elmegreen16,elmegreen18}, revealed a $2.9\times10^7\;M_\odot$ molecular cloud at the hotspot (assuming $\alpha_{\rm CO}=100\;M_\odot$ (K km s$^{-1}$ pc$^2$)$^{-1}$ for  14\% solar metallicity).

\begin{summary}[SUMMARY POINTS] \label{sect-summary}
\begin{enumerate}
\item Dwarf irregular galaxies are gas-rich, low-density, metal-poor, and forming stars. 
As such, they are important stress-tests for models of star formation.
\item Dwarf irregulars are thicker than spirals and this affects star formation by making disk self-gravity weaker and 3D processes more important.
\item Stellar feedback is an important process in the evolution of dwarf galaxies because it can form large shells that may trigger star formation, dump metals into the CGM, and increase the porosity of the gas, and because
it can significantly modify the disk gravitational potential, causing stars to scatter around.
\item Molecular cloud structure in dwarf galaxies consists of tiny CO cores in thick PDRs. 
However, the CO cores are similar to those in the Milky Way in terms of density, size--line-width relation, and extinction.
There could be substantial gas in dwarfs not traced by CO or \hi.
\item The dust-to-gas ratios are lower in dwarfs compared to spirals, the dust equilibrium temperatures are warmer, the dust grain sizes are smaller, and PAHs are nearly missing.
\item The low metallicities and dust-to-gas ratios in dIrrs affect the ISM and star formation by decreasing the mass fraction of cold molecular gas.
\end{enumerate}
\end{summary}

\begin{issues}[FUTURE ISSUES] \label{sect-future}
\begin{enumerate}
\item Do we see evolutionary consequences of the large-scale dynamics predicted by dwarf galaxy simulations?

\item What drives turbulence in dwarf galaxies? New observations and tests are needed.

\item How are the self-gravitating \HI\ clouds that are precursors to star-forming molecular clouds formed? 
High angular and spectral resolution radio interferometer surveys may be able to address this. 

\item Are dwarfs a significant source of reionization in the early universe? High-angular-resolution, high-sensitivity FIR observations may be able to answer this.

\item What are the consequences of the different structure of molecular clouds on the star formation process and resulting stars? High-resolution and high-sensitivity radio interferometers are pushing maps of molecular clouds to lower metallicities, but at some metallicity will there be no CO in star-forming regions? How can we better trace CO-dark gas and what is its nature, amount, and distribution?

\item Where is star formation taking place within low-metallicity star-forming regions? Can it occur in dark H$_2$ or cold \HI\ without CO?
In what component of the star-forming region are clumps of massive stars formed?
How does this local, low-metallicity star formation compare with star formation at high redshift?
High resolution, high-sensitivity FIR observations may answer these questions.

\item What is the reason for the dearth of PAH features in dwarf galaxies? How does the dust-to-gas mass ratio behave at extremely low metallicities?
How has the dust composition in dIrrs evolved over time? 

\item What is the source of the excess submm emission?

\item What is the nature of the CGM around dwarf irregulars and how do the CGM and IGM affect the 
evolution of these tiny galaxies? Spectra of background sources 
and ultra-deep narrow-band imaging  
may address this.

\item How far does the gas disk extend in dwarf irregulars and what role does highly extended gas play in the evolution
of the galaxy? What fraction of dwarfs have evidence for cold-accretion from the IGM? 
Radio interferometers that can trace \HI\ to very low column densities may be able to demonstrate how the IGM connects to the rest of the galaxy.
\end{enumerate}
\end{issues}

\section*{DISCLOSURE STATEMENT}
The authors are not aware of any affiliations, memberships, funding, or financial holdings that
might be perceived as affecting the objectivity of this review.

\section*{ACKNOWLEDGMENTS}
We are indebted to the referee, Nia Imara, and to Maarten Baes, Shmuel Bialy, Francoise Combes, Fr\'{e}d\'{e}ric Galliano, John Gallagher III, Leslie Hunt,
Monica Rubio, and Mark Wolfire for helpful suggestions to improve this chapter.
DAH is grateful for funding from the National Science Foundation through AST-1907492.
Lowell Observatory sits at the base of mountains sacred to Native tribes throughout the region.
We honor their past, present, and future generations, who have lived here for millennia 
and will forever call this place home.

%


\begin{table}[h]
\tabcolsep7.5pt
\caption{Major surveys of the ISM of dwarf irregular galaxies}
\label{tab1}
\begin{center}
\begin{tabular}{@{}l|l|c|c|c|c|l@{}}
\hline 
            &                 &                 & Angular res & Spectral res & Distance &                  \\
Survey &Telescope--Spec region & \# Dwfs &      {(}arcsec) &    {(}\kms )       &   {(}Mpc)   & Ref \\
\hline
\HI\ Survey            & Arecibo-\HI            & 123  & 198  & 6,12,24  & 6.4--123    & 1 \\
FIGGS                & GMRT-\HI               & 65   & 5,22 & 1.65       & 1.0--17.4   & 2 \\
THINGS              & VLA-\HI                  & 12   & 6      & 1.3--5.2  & 2.0--5.3     & 3 \\
ALFALFA             & Arecibo-\HI             & 229 & 210  & 5            & 0--49         & 4 \\
LITTLE THINGS  & VLA-\HI                  & 41   & 6    & 1.3, 2.6    & 0.7--10.3   & 5 \\
VLA-ANGST      & VLA-\HI                    & 35   & 6    & 0.6--2.6    & 1.3--4.0      & 6 \\
DGS                  & {\it Herschel}-FIR      & 50   &  6--36     &  150--200$^a$   & 0.7--191    & 7 \\
ATLAS$^{3D}$-MATLAS & WSRT, Arecibo  & 59 & 60, 240 & 16, 10 & 2--99  & 8 \\
LGLBS               & VLA-\HI, continuum  &  4   &   6  & 0.42        &  0.5--1.0    & 9 \\
\hline
\end{tabular}
\end{center}
\begin{tabnote}
1 -- \citet{Salzer:02}; 2 -- \citet{Begum:08}; 3 -- \citet{Walter:08}; 4 -- \citet{Huang:12};
5 -- \citet{Hunter:12}; 6 -- \citet{Ott:12}; 7 -- \citet{madden13}; 
8 -- \citet{Poulain:22}; 9 -- \cite{koch24}
\end{tabnote}
$^a${Velocities are related to the PACS spectral survey. 
The PACS and SPIRE photometric survey resolving power is $\sim$2 to 3.}
\end{table}

\end{document}